\UseRawInputEncoding

\documentclass[aps,pre,amsmath,amssymb,floatfix,reprint,citeautoscript,noeprint,superscriptaddress,twocolumn]{revtex4-2}
\usepackage{lineno}

\usepackage{dsfont}
\usepackage{lipsum} 
\usepackage{bibentry}
\usepackage[english]{babel}
\usepackage[dvipsnames]{xcolor}
\usepackage{graphicx}
\usepackage[caption=false]{subfig} 
\usepackage{amsmath,amssymb,bm}
\usepackage[version=3]{mhchem}
\usepackage{verbatim}
\usepackage{multirow}
\usepackage{dcolumn}
\usepackage{float}
\usepackage{nicefrac}
\usepackage{siunitx}
\usepackage{booktabs}
\usepackage{chemformula}
\usepackage{wrapfig}
\usepackage{enumitem} 
\setlist[itemize]{leftmargin=1.5em, labelsep=0.5em}
\usepackage{transparent}
\usepackage{xr-hyper}                 
\usepackage[colorlinks,allcolors=black,citecolor=blue,urlcolor=blue]{hyperref}
\usepackage{tcolorbox}
\usepackage{relsize}
\hypersetup{
    citecolor=Turquoise,
    colorlinks=true,
    linkcolor=black,    
    urlcolor=Turquoise,
    }
\usepackage{titlesec}
    
\DeclareSIUnit[number-unit-product = {\,}]{\amu}{amu}
\DeclareSIUnit[number-unit-product = {\,}]{\kJmol}{\kilo\joule\per\mol}
\DeclareSIUnit[number-unit-product = {\,}]{\Nsm}{\newton\second\per\meter\cubed}
\DeclareSIUnit[number-unit-product = {\,}]{\THz}{\tera\hertz}
\DeclareSIUnit[number-unit-product = {\,}]{\meV}{\milli\electronvolt}
\DeclareSIUnit[number-unit-product = {\,}]{\cal}{cal}

\newcommand{\tbf}[1]{\textbf{#1}}

\newcommand{\mbf}[1]{\boldsymbol{\mathit{#1}}}
\newcommand{\mrm}[1]{\mathrm{#1}}
\newcommand{\mcal}[1]{\mathcal{#1}}

\usepackage{sansmathfonts}
\usepackage[justification=centerlast, font={sf}, labelfont={bf}]{caption}
\begin{document}

\title{Characterizing adsorption thermodynamics in carbon capture materials: an efficient first-principles approach}

\author{Anna T. Bui}
\email{annatbui@stanford.edu}
\affiliation{Department of Chemistry, Stanford University, Stanford, California, 94305, USA}

\author{Grant M. Rotskoff}
\email{rotskoff@stanford.edu}
\affiliation{Department of Chemistry, Stanford University, Stanford, California, 94305, USA}

\author{Thomas E. Markland}
\email{tmarkland@stanford.edu}
\affiliation{Department of Chemistry, Stanford University, Stanford, California, 94305, USA}

\date{\today}

\begin{abstract}
Porous materials offer a vast design space for CO$_2$ capture, whose exploration hinges on efficient and accurate first-principles predictions of the thermodynamics governing adsorption. However, simulating isotherms and heats of adsorption using grand canonical molecular simulations is extremely computationally expensive when combined with accurate ab initio treatments of the molecular interactions, limiting screening to low-pressure uptake encoded in the Henry constant. Identifying and designing the best CO$_2$ capture materials requires a strategy that can efficiently calculate their adsorption properties using ab initio potential energy surfaces such as those now available from foundational machine-learned interatomic potentials (MLIPs). Here we develop a neural classical density functional (cDFT) framework for carbon capture processes that allows direct calculation of adsorption isotherms from low to high loading limits of metal-organic frameworks (MOFs) without grand canonical simulations. We demonstrate the success of our approach by calculating the ab initio adsorption isotherms for over 5,000 experimentally reported MOFs. From our screening, we show how these materials can be characterized using four isotherm shapes, arising from their pore size and binding sites, which the Henry constant cannot distinguish. We show that these shapes, rather than the binding strength alone, decide which MOFs suit which process: small pores with strong binding sites for vacuum and temperature swings at low pressure, large pores with weakly binding walls for pressure swings. Finally, we identify the most promising candidates by weighing their working capacity against the energy needed for regeneration.
\end{abstract}

\maketitle

\section{Introduction}

Anthropogenic emissions of carbon dioxide (CO$_2$) are the main driver of global climate change, one of the most pressing challenges facing society \cite{IPCC2023AR6SYR}. Even with rapid cuts in emissions, limiting global warming will require billions of tonnes of CO$_2$ to be captured each year \cite{Smith2024StateCDR}. Both point-source and direct air capture call for materials that bind CO$_2$ strongly enough to separate it, yet weakly enough to release it again at low energy cost, typically by lowering the pressure, raising the temperature or both, in a swing adsorption process \cite{farmahini2021performancebased}. Metal--organic frameworks (MOFs) are leading candidates for this task, owing to their high porosity and the near-endless chemical tunability of their metal nodes and organic linkers \cite{Trickett2017MOFCO2, zeng2022porous}. This very tunability, however, also presents a challenge: with tens of thousands of synthesized frameworks \cite{Zhao2025CoREMOFDB}, and hundreds of thousands of hypothetical ones \cite{wilmer2012largescale, boyd2019datadriven}, identifying the best material for each swing under its operating conditions by experiment alone is impractical. Computational screening has therefore become an indispensable tool for materials discovery \cite{Moghadam2024ComputationalDiscovery, colon2014highthroughput, Ongari2020TooManyMaterials, Boyd2017NanoporousGenome, wilmer2012largescale, wilmer2012structureproperty, chung2016silico, Rosen2021QMOF}, but its success hinges on the ability to predict adsorption properties efficiently from first principles.

At its heart, adsorption is governed by the thermodynamic balance between the CO$_2$ inside the framework material and the surrounding gas, which share the same chemical potential---set by the temperature and pressure. A key quantity that governs the performance of MOFs is therefore the adsorption isotherm, which measures the CO$_2$ uptake as a function of pressure at a given temperature. Predicting it faithfully requires an accurate description of two main ingredients: the thermodynamics of CO$_2$'s vapor--liquid equilibria, and the interaction of CO$_2$ with the solid framework. While the former can be well described by empirical force fields \cite{Harris1995EPM2, Potoff2001VaporLiquid}, the latter remains extremely challenging: the chemical space of MOFs is vast, and interactions with moieties such as open metal sites demand quantum mechanical accuracy. Foundational machine-learned interatomic potentials (MLIPs) \cite{Wood2025UMA, Batatia2025Foundation, Rhodes2025OrbV3}, trained on large quantum mechanical datasets, can offer such accuracy across the periodic table at a fraction of the cost of electronic structure calculations, and have recently been tailored to carbon capture in MOFs \cite{sriram2025odac25, Lim2025DirectAirCapture, Inizan2026uMOF}. However, accurately predicting these models' adsorption properties remains an open challenge.

At low pressure, the uptake grows linearly with pressure, following Henry's law, $n = K_\mrm{H}P$, and is set by the Henry constant $K_\mrm{H}$ alone. At higher pressure, however, the isotherm becomes far more complex: as strong binding sites saturate and the pores fill, the uptake grows more slowly than the pressure, whereas CO$_2$--CO$_2$ attraction can make it grow faster. Process-relevant metrics, such as working capacities \cite{rajagopalan2016adsorbent, raganati2020capture}, therefore depend on the full shape of the isotherm, which $K_\mrm{H}$ alone cannot capture. Obtaining a full adsorption isotherm requires extensive grand canonical Monte Carlo (GCMC) simulations \cite{Goeminne2023Adsorption, Wang2026FrameworkFlexibility, Edwards2026MLIPMC, Charalambous2024CarbonCapture}: even with empirical force fields, screening studies simulate only a few pressures \cite{wilmer2012largescale, boyd2019datadriven}, and curated databases of simulated isotherms cover only hundreds of MOFs \cite{oliveira2023crafted}. With MLIPs, the cost is higher still by orders of magnitude: full isotherms, i.e., those spanning the dilute to saturation limits, have been computed for only a handful of MOFs \cite{Goeminne2023Adsorption, Edwards2026MLIPMC, Inizan2026uMOF}, and screening across many MOFs has been restricted to the dilute limit \cite{Lim2025DirectAirCapture}.

Classical density functional theory (cDFT) \cite{Evans1979NatureLiquidVapour, Hansen2013SimpleLiquids} is an exact statistical mechanical framework for inhomogeneous fluids that provides direct access to the chemical potential and the rich thermodynamic properties that stem from it, e.g., adsorption isotherms and heats of adsorption, thus obviating the need for GCMC simulations. In cDFT, the equilibrium density $\rho(\mbf{r})$ and thermodynamics of a fluid at temperature $T$ follow from its excess free energy functional $\mcal{F}_\mrm{ex}([\rho],T)$, which arises from intermolecular interactions \cite{Mermin1965Thermal, Evans1979NatureLiquidVapour, Hansen2013SimpleLiquids}. In practice, however, $\mcal{F}_\mrm{ex}$ is generally unknown; accurate approximations are well established only for simple fluids such as hard spheres \cite{Rosenfeld1989FMT}, and applications to adsorption have largely relied on simple models of the fluid \cite{Granderath2026ProcessScreening, DufourDecieux2025Adsorption}. In recent years, machine learning (ML) has overcome this obstacle: ML functionals learned directly from simulation data \cite{Sammuller2023NeuralFunctional} have described liquid--gas coexistence \cite{Sammuller2025LiquidGas}, mixtures \cite{Robitschko2025LiquidLiquid}, electrolytes and molecular fluids \cite{Bui2025IonicFluids, Yang2025OperatorLearning, Bui2026Dielectrocapillarity} and, when combined with MLIPs, liquids from first principles \cite{Bui2026MultiscaleLiquids, Chen2026PhaseField}. While most of these functionals have been learned from density profiles in planar geometry, they have more recently been extended to three-dimensional density fields~\cite{Cheng2026Equivariant, Weimar2026CubicEquivariant, Chen2026PhaseField}.

Here, we introduce a framework for efficient ab initio calculation of carbon capture thermodynamics by combining a neural functional of the CO$_2$ fluid with electronic potential energy surfaces from foundational MLIPs. We show that this allows us to obtain the full adsorption isotherms, from low to high loading, of MOFs without grand canonical simulations, at a cost set by just a single grid of interaction energies per MOF. Our approach is particularly well suited to high-throughput screening of materials for CO$_2$ adsorption from first principles: for a given fluid, $\mcal{F}_\mrm{ex}$ is unique and independent of the framework, which enters only through the external potential, so that the functional is learned once and transfers to every MOF. Validating eight MOF--CO$_2$ potentials against experimental isotherms of 20 well-characterized MOFs, we find that, among the MLIPs, only UMA-m~\cite{Wood2025UMA} and the MOF-specific uMOF-MH~\cite{Inizan2026uMOF} closely reproduce the measured uptake. Applying our approach to over 5,000 experimentally reported frameworks of the CoRE MOF 2025 database~\cite{Zhao2025CoREMOFDB}, we find that their isotherms fall into four characteristic shapes, which the Henry constant alone cannot distinguish, and we trace each, through the computed CO$_2$ density, to its origin in the pore size and the binding sites. We then show that it is the isotherm's shape, rather than the binding strength alone, that decides which MOFs are optimal for a given swing: steep Langmuir MOFs, with strong sites in small pores, excel at vacuum and temperature swings at low pressure; S-shaped and two-step MOFs, with large pores, at pressure swings; while the gradual Langmuir isotherms of most MOFs rarely rank among the best. Finally, accounting for the regeneration energy, we identify the top candidates for each swing condition, and the chemical and structural motifs needed to achieve them.

\section{Neural cDFT: accurate and efficient isotherms without grand canonical sampling}

\begin{figure}[t]
  \includegraphics[width=\linewidth]{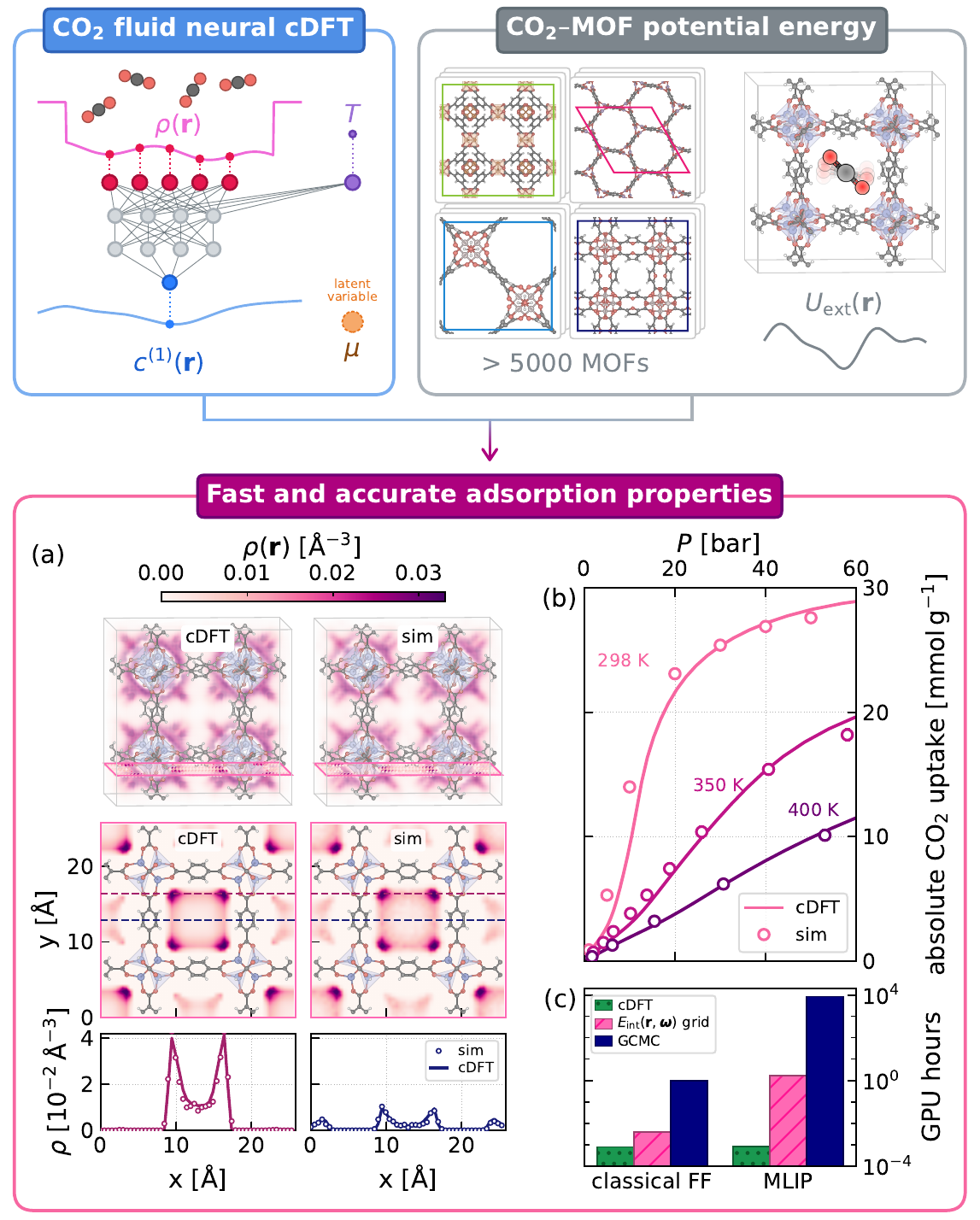}
  \caption{\tbf{Adsorption thermodynamics from a learned CO$_2$ fluid functional and a CO$_2$--MOF potential.} The neural cDFT maps density $\rho(\mathbf{r})$ and temperature $T$ to the one-body direct correlation function $c^{(1)}(\mathbf{r})$, with one latent chemical potential $\mu$ per training simulation (top left). Each MOF's external potential $U_\mrm{ext}(\mathbf{r})$ is obtained from the rigid CO$_2$ interaction energy $E_\mrm{int}(\mbf{r},\bm{\omega})$ sampled over positions and orientations (top right). For IRMOF-1, cDFT and simulations using TraPPE for CO$_2$ and UFF for the framework show excellent agreement in (a) the CO$_2$ density at 298\,K and about 10\,bar (cDFT, left, and MD, right) and (b) the absolute uptake at 298, 350 and 400\,K (cDFT, lines, and GEMC \cite{Borycz2016CO2Adsorption} and GCMC, circles).  (c) Even including the interaction energy grid calculation, cDFT isotherms take less GPU wall time than GCMC for both the classical force field (UFF) and the MLIP (UMA-m)).}
  \label{fig1}
\end{figure}

Our framework to efficiently obtain first-principles adsorption thermodynamics without grand canonical sampling using neural cDFT is outlined in Figure~\ref{fig1}. In this section, we explain how we construct a neural cDFT functional for the CO$_2$ fluid that is transferable across a wide range of porous systems; we explain how we can combine it with CO$_2$--MOF external potentials obtained from MLIPs and then show how this can be used to compute the required thermodynamics of adsorption that can be compared to experimentally measured quantities.

\textit{Constructing neural cDFT functional of CO$_2$.} The excess free energy functional is a property of the CO$_2$ fluid alone, i.e., it is independent of any external potential. Once it is known, cDFT yields the equilibrium structure and thermodynamics of the fluid in any framework. At chemical potential $\mu$ and temperature $T$, in the external potential $U_\mrm{ext}(\mbf{r})$ that the framework exerts on CO$_2$, the equilibrium density satisfies
   \begin{equation}
  \label{eqn:EL}
  \rho(\mbf{r}) = \frac{\zeta}{\Lambda^{3}}\exp\big[- \beta\big(U_{\mathrm{ext}}(\mbf{r}) - \mu\big) + c^{(1)}(\mbf{r};[\rho],T)\big],
\end{equation}
where $\Lambda$ is the thermal de Broglie wavelength and $\zeta$ is an intramolecular partition function, needed because $\mu$ couples to the number of molecules rather than atoms. The only unknown in Eq.~\eqref{eqn:EL} is the one-body direct correlation functional $c^{(1)}$, which we represent with a neural network, in a development dubbed ``neural cDFT'' \cite{Sammuller2023NeuralFunctional, Bui2026MultiscaleLiquids}. Crucially, $c^{(1)}$ is short-ranged, such that its value at $\mbf{r}$ is determined by the density within a cutoff of $\mbf{r}$ (Fig.~\ref{fig1}, top left). The network therefore takes as input the local density around each point, together with the temperature, and returns $c^{(1)}(\mbf{r})$. We generated the training data from canonical molecular dynamics of CO$_2$ in random 3D external potentials at approximately 220--500\,K, supplemented by simulations of the bulk fluid; canonical sampling avoids the particle insertions that make grand canonical simulations impractical with MLIPs. As the chemical potential of each training simulation is not known a priori, we treated it as a latent variable and learned it together with $c^{(1)}$ by using Eq.~\eqref{eqn:EL} itself as the loss function \cite{Sammuller2026ChemicalPotential, Bui2026MultiscaleLiquids}. This approach has previously been demonstrated for fluids in planar geometry \cite{Bui2026MultiscaleLiquids}. Here, since adsorption in a MOF is intrinsically three-dimensional, we extend neural cDFT to resolve the density fields on a 3D grid. We trained functionals both for the TraPPE force field \cite{Potoff2001VaporLiquid} and at the PBE-D3 level of theory \cite{Mathur2023CO2PhaseBehavior}. To efficiently train the PBE-D3 cDFT functional we performed fine-tuning from the TraPPE cDFT functional using simulations driven by an MLIP. The PBE-D3 functional thus amounts to a genuinely \textit{ab initio} cDFT for CO$_2$. Once trained, the functional can be transferred to any porous material without retraining since the effect of the material enters only through the external potential, $U_\mrm{ext}$.

\textit{Computing external potentials of MOFs from foundational MLIPs.} For a given MOF, we compute the interaction energy $E_{\mrm{int}}(\mbf{r}, \bm{\omega})=E_{\mrm{MOF+CO}_2}(\mbf{r},\bm{\omega}) - E_{\mrm{MOF}} - E_{\mrm{CO}_2}$ of a single, rigid CO$_2$ molecule in a rigid MOF, on a grid of positions $\mbf{r}$ and orientations $\bm{\omega}$, at the level of theory of the chosen interatomic potential, each requiring a single-point energy calculation. To couple it to the number density, at the temperature $T$ of interest, we Boltzmann-average over CO$_2$ orientations to obtain the external potential
\begin{equation}
U_{\mathrm{ext}}(\mbf{r})
  = -k_{\mrm{B}}T
    \ln \left\langle \mrm{exp} \left({-\beta E_{\mrm{int}}(\mbf{r},\bm{\omega})} \right) \right\rangle_{\bm{\omega}}.
    \label{eq:effective-potential}
\end{equation}
This average is exact in the dilute limit and remains accurate at finite loading, except in the few MOFs whose binding sites restrict CO$_2$ to a narrow range of orientations, as shown in Fig.~S3.
This information alone is sufficient to obtain the Henry constant,
\begin{equation}
K_{\mrm{H}}=\frac{\beta}{V_{\text{cell}}}\int\!\!\mrm{d}\mbf{r}\,
    \mrm{exp}(-\beta U_{\mrm{ext}}(\mbf{r})),
    \label{eq:kh}
\end{equation}
and the zero-uptake isosteric heat of adsorption $Q^0_\mrm{st}$,
\begin{equation}
Q^0_\mrm{st} = k_\mrm{B}T - \frac{\int\!\mrm{d}\mbf{r}\,\left\langle E_{\mrm{int}}(\mbf{r},\bm{\omega})\,\mrm{exp}\left(-\beta E_{\mrm{int}}(\mbf{r},\bm{\omega})\right)\right\rangle_{\bm{\omega}}}
    {\int\!\mrm{d}\mbf{r}\,\left\langle \mrm{exp}\left(-\beta E_{\mrm{int}}(\mbf{r},\bm{\omega})\right)\right\rangle_{\bm{\omega}}}.
    \label{eq:qst0}
\end{equation}

\textit{Adsorption isotherms from cDFT.} 
In an adsorption measurement, the MOF is held in contact with a reservoir of CO$_2$ gas at fixed $T$ and $P$, and CO$_2$ enters the pores until the adsorbed fluid and the gas share the same chemical potential $\mu$. In cDFT, the reservoir pressure follows from the bulk fluid at the same $\mu$, whose density $\rho_\mrm{b}$ solves $\beta\mu = \ln(\zeta^{-1}\Lambda^3\rho_\mrm{b}) - c^{(1)}([\rho_\mrm{b}],T)$, i.e.,
\begin{equation}
  \beta P = \rho_\mrm{b}\left[1 - c^{(1)}([\rho_\mrm{b}],T)\right]
            + \rho_\mrm{b}\int_0^1\!\!\mrm{d}\alpha\; c^{(1)}([\alpha\rho_\mrm{b}], T).
  \label{eq:pressure}
\end{equation}
At each $\mu$, we solve Eq.~\eqref{eqn:EL} self-consistently. From this we obtain the absolute uptake,
\begin{equation}
    n(\mu) = \frac{1}{M}\int\!\mrm{d}\mbf{r}\, \rho(\mathbf r; \mu),
\end{equation}
where $M$ is the framework mass and the isotherm $n(P)$ follows by mapping each $\mu$ to its $P$ through Eq.~\eqref{eq:pressure}. For comparison with experiment, which measures the excess uptake, we subtract the bulk gas that would occupy the accessible volume $V_\mrm{acc}$, $n_\mrm{ex} = n - \rho_\mrm{b}V_\mrm{acc}/M$.

\textit{Comparison of cDFT to grand canonical simulations.} 
Figure~\ref{fig1}(a) compares the 3D CO$_2$ density obtained from molecular dynamics simulations and computed from cDFT for IRMOF-1 \cite{li1999mof5} at the same uptake. This MOF has only a few hundred atoms, and we used a classical force field to ensure that we could perform enough sampling to fully converge the simulations. Our cDFT results are in near perfect agreement with our molecular dynamics simulations for both the CO$_2$ 3D density and the two density cuts shown in Fig.~\ref{fig1}(a).

Isotherms from Gibbs ensemble Monte Carlo (GEMC) \cite{Borycz2016CO2Adsorption} at 298~K and from our grand canonical Monte Carlo (GCMC) simulations at 350~K and 400~K (circles) are compared with our cDFT results (lines) in Fig.~\ref{fig1}(b). Again, excellent agreement is obtained. Fig.~S5 also shows that in the dilute limit our cDFT results for the Henry constant, $K_\mrm{H}$, and zero-uptake heat of adsorption $Q^0_\mrm{st}$ agree to within 1\,kJ\,mol$^{-1}$ of values recently obtained using Widom insertion with MOF--CO$_2$ MLIPs~\cite{Lim2025DirectAirCapture, sriram2025odac25}. However, as we show in Sec.~\ref{sec:theory_vs_expt}, since our cDFT approach provides access to the full loading curve, it lets us characterize important MOF adsorption properties beyond the Henry regime. 

With an MLIP, the main cost is the $E_\mrm{int}(\mbf{r},\bm{\omega})$ grid, which takes up to approximately two hours for UMA-m on an NVIDIA H100 GPU but is needed only once per MOF (Figs.~\ref{fig1}(c) and S6).  Given this grid, the cDFT isotherm then takes seconds, since it solves directly for the CO$_2$ density. GCMC, in contrast, must sample configurations of many interacting CO$_2$ molecules at every pressure. With the MLIP evaluated at every configuration, this is impractical for more than a few MOFs. Even when the MOF--CO$_2$ interactions are evaluated on a grid and only the CO$_2$--CO$_2$ interactions are computed explicitly, the sampling remains more costly than solving for the density, and increasingly so as the pores fill.

\section{Comparison of calculated and experimental adsorption isotherms from the dilute limit to saturation} \label{sec:theory_vs_expt}

To assess the predictive power of foundational MLIPs for adsorption, we first compare our cDFT framework to experiments for the 20 well-known MOFs at room temperature. Experimental isotherms extending to saturation are very limited, and hence we compiled the available data from multiple literature sources \cite{Millward2005CarbonDioxideStorage, Charalambous2024CarbonCapture, cavka2014high}. These MOFs span the main families studied for CO$_2$ capture (Fig.~S7): Zn$_4$O frameworks such as IRMOF-1 and MOF-177 \cite{chae2004mof177}, Zr frameworks such as UiO-66 \cite{cavka2008uio66}, the Cu paddlewheel HKUST-1 \cite{chui1999hkust}, MOF-74 analogs with open metal sites \cite{rosi2005mof74, caskey2008mof74}, Al frameworks such as MIL-160 and small-pore physisorbents such as CALF-20 \cite{lin2021scalable}.

\begin{figure}[t]
  \includegraphics[width=\linewidth]{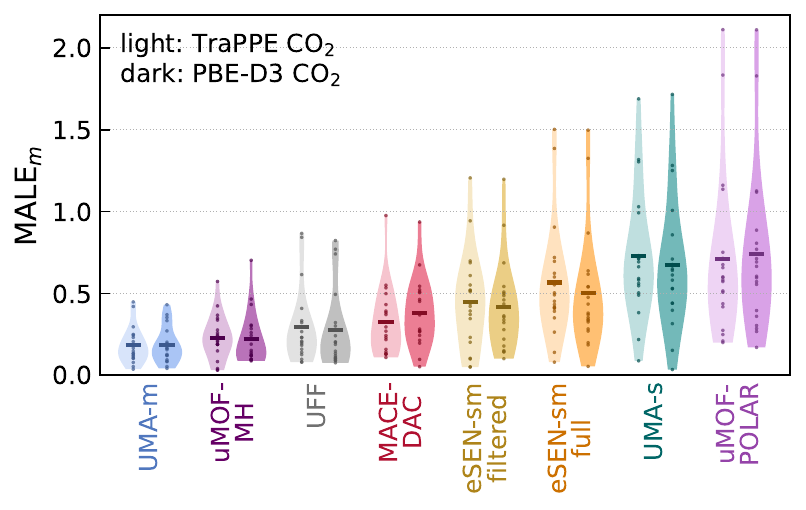}
\caption{\tbf{Assessment of MLIP accuracy in predicting experimentally measured CO$_2$ excess uptake.} Dots give the mean absolute log error $\mrm{MALE}_m$ for each of the 20 MOFs in Fig.~S7 at 298\,K. For each MOF--CO$_2$ potential and CO$_2$ functional (TraPPE, light or PBE-D3, dark), violins show the error distribution and bars mark its mean (MALE).}
\label{figmale}
\end{figure}

Our approach allows us to directly test whether current-generation foundational MLIPs can accurately capture the experimental adsorption isotherms. For the MOF--CO$_2$ interaction, we consider UFF \cite{Rappe1992UFF}, a standard classical force field for MOFs; the uMOF models, uMOF-MH and uMOF-POLAR, fine-tuned from the MACE-MH-1 and polarizable MACE-POLAR-1 foundation models on r$^2$SCAN-D4 calculations of MOFs, including frameworks loaded with CO$_2$ and other gases \cite{Inizan2026uMOF}; MACE-DAC, fine-tuned for the interactions of CO$_2$ and water with MOFs \cite{Lim2025DirectAirCapture}; the eSEN models trained on the full and filtered ODAC25 datasets~\cite{Fu2025eSEN, sriram2025odac25}; and the UMA small and medium models (UMA-s and UMA-m, both v1.1) \cite{Wood2025UMA}, whose training data contain the ODAC25 subset together with other materials. To assess the accuracy, Fig.~\ref{figmale} shows the distribution of the mean absolute log error (MALE) across the 20 MOFs for each MLIP when combined with our CO$_2$ cDFT functionals obtained from TraPPE and PBE-D3.

\begin{figure*}[t]
  \includegraphics[width=\linewidth]{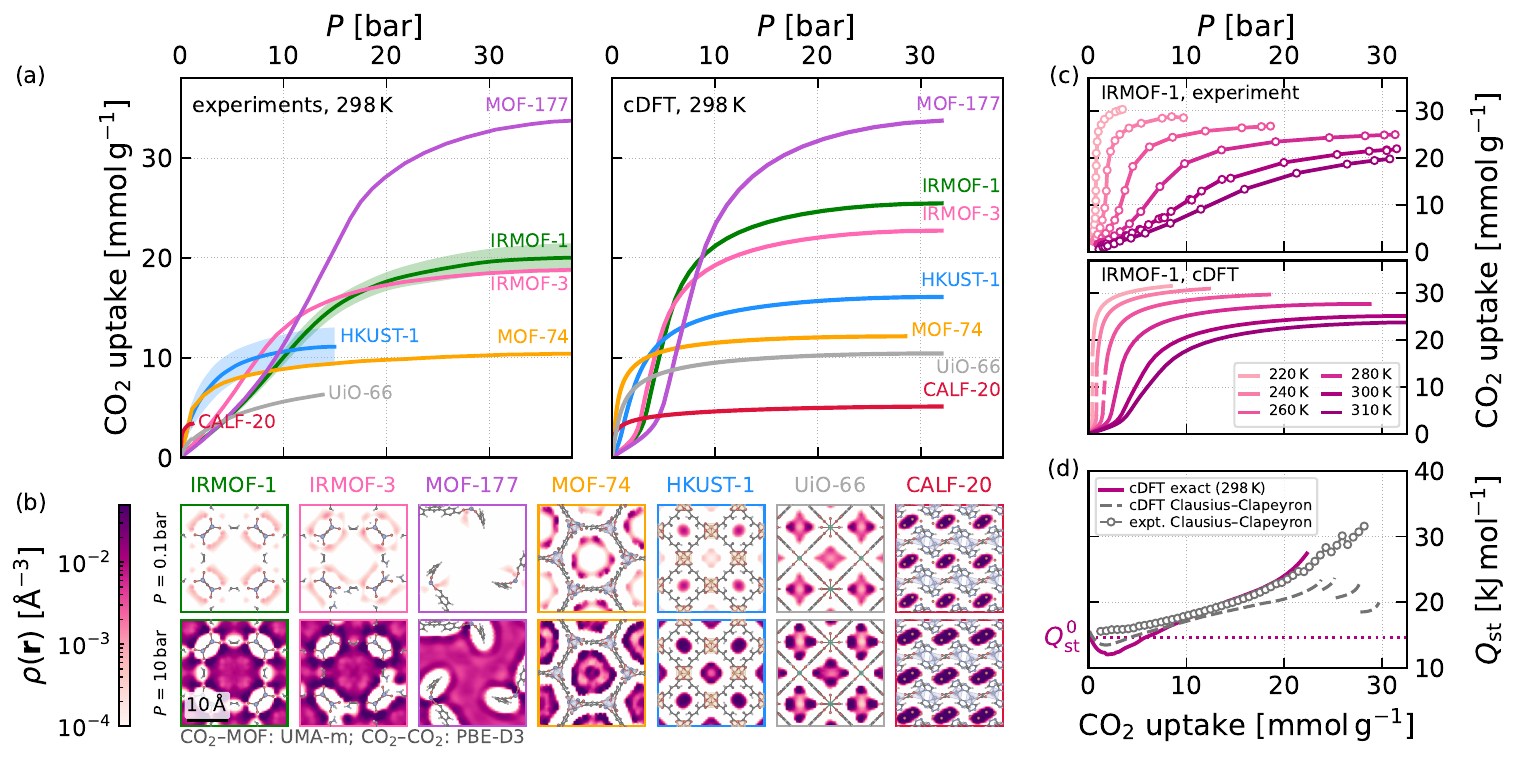}
  \caption{\tbf{Comparison of cDFT and experiment for CO$_2$ adsorption isotherms and isosteric heat.} For seven MOFs from Fig.~S7, (a) compares mean experimental excess uptake isotherms at 298\,K (left) with cDFT predictions (right) using UMA-m and the PBE-D3 CO$_2$ functional. The corresponding CO$_2$ density maps in (b) show the plane through the minimum of $U_\mrm{ext}$ in each MOF at 0.1\,bar (top) and 10\,bar (bottom). For IRMOF-1, (c) compares experimental \cite{simmons2011carbon} and cDFT isotherms (top and bottom, respectively) from 220 to 310\,K. In (d), the exact cDFT isosteric heat $Q_\mrm{st}(n)$ at 298\,K (solid) is calculated from the fugacity and compared with Clausius--Clapeyron estimates from cDFT (dashed) and experiment (circles). The dotted line marks the zero-uptake limit $Q^0_\mrm{st}$.}
  \label{fig2}
\end{figure*}

UMA-m is the best-performing foundational MLIP, with a MALE of 0.19, i.e., on average its isotherms agree with experiment to within a factor of $\sim$1.5. It is closely followed by uMOF-MH, with a MALE of 0.22, at an order of magnitude lower computational cost. For IRMOF-1, UMA-m deviates from experiment with a MALE of 0.20, three times the 0.07 between cDFT and GCMC for the same MOF (Fig.~\ref{fig1}(b)). The largest error relative to experiment therefore arises from the MOF--CO$_2$ interactions and not from our cDFT approach. 

Notably, the accuracies are relatively insensitive to whether the functional was trained on TraPPE or PBE-D3 CO$_2$--CO$_2$ interactions (light vs dark shading). The full data and correlation plots are shown in Figs.~S8 and S9, which allow one to easily identify the cases where there is systematic over- or underestimation of uptake. UFF appears competitive with a MALE of 0.28, but mainly because most of the 20 MOFs lack strong binding sites. Where binding is strong, as in Mn-MOF-74, CALF-20 and NOTT-300, it underestimates the uptake by factors of 5--7 (Table~S1). 

In Fig.~\ref{fig2}(a), we show the full isotherms for seven representative MOFs. Our approach captures not only the relative uptake of the MOFs, but also the shapes of their isotherms: the first three are S-shaped, filling mostly at high pressure, while the other four are Langmuir-like, filling rapidly at low pressure (we return to isotherm shapes in Sec.~\ref{sec:classification}). Moreover, cDFT provides direct access to the microscopic CO$_2$ densities within the materials that give rise to their adsorption which are shown at low (0.1\,bar) and high (10\,bar) pressures in Fig.~\ref{fig2}(b).

Our approach also faithfully captures how adsorption changes with temperature: the computed isotherms of IRMOF-1 follow those experimentally measured from 220 to 310\,K \cite{simmons2011carbon}, as shown in Fig.~\ref{fig2}(c). Isotherms at several temperatures can be used to obtain the isosteric heat of adsorption, the heat released as CO$_2$ adsorbs at loading $n$, through the Clausius--Clapeyron relation,
\begin{equation}
      Q_\mrm{st}(n) \simeq -R\left(\frac{\partial \ln P}{\partial (1/T)}\right)_{n}.
    \label{eq:qst}
\end{equation}
Experimentally, Eq.~\eqref{eq:qst} is usually used with isotherms at two or a few temperatures, which assumes an ideal gas and a temperature-independent $Q_\mrm{st}$ \cite{nuhnen2020practical}. 
Since isotherms are usually measured only at room temperature \cite{simmons2011carbon, Mason2011PostcombustionTSA}, our approach provides a way to compute them over a wide range of temperatures. $Q_\mrm{st}(n)$ then follows at any temperature from closely spaced isotherms, without the ideal-gas approximation, since the functional provides the bulk equation of state of CO$_2$.

Fig.~\ref{fig2}(d) compares the computed and experimental~\cite{simmons2011carbon} isosteric heat of adsorption, $Q_\mrm{st}(n)$, for IRMOF-1. Because cDFT can give $Q_\mrm{st}(n)$ at each temperature, we also show how it varies from 220 to 298\,K in Fig.~S10. In IRMOF-1, $Q_{\rm st}(n)$ first dips slightly as its strongest binding sites, near the Zn$_4$O clusters, saturate, and then increases by 15~kJ\,mol$^{-1}$ as the pore fills, nearly doubling. This is consistent with the rise seen in experiment, and reflects that once CO$_2$ fills the pore, it binds to other CO$_2$ molecules as much as to the framework. The zero-loading value $Q^0_\mrm{st}$, which is usually used for screening, sees only the framework and misses this rise from CO$_2$--CO$_2$ interactions, which our cDFT calculations fully capture. As we will show in Sec.~\ref{sec:classification}, these CO$_2$--CO$_2$ attractions drive the S-shaped isotherms obtained for many MOFs in the database.

\section{High-throughput screening and characterization of isotherm shapes in the CoRE MOF database}
\label{sec:classification}

\begin{figure*}[t]
  \includegraphics[width=0.9\linewidth]{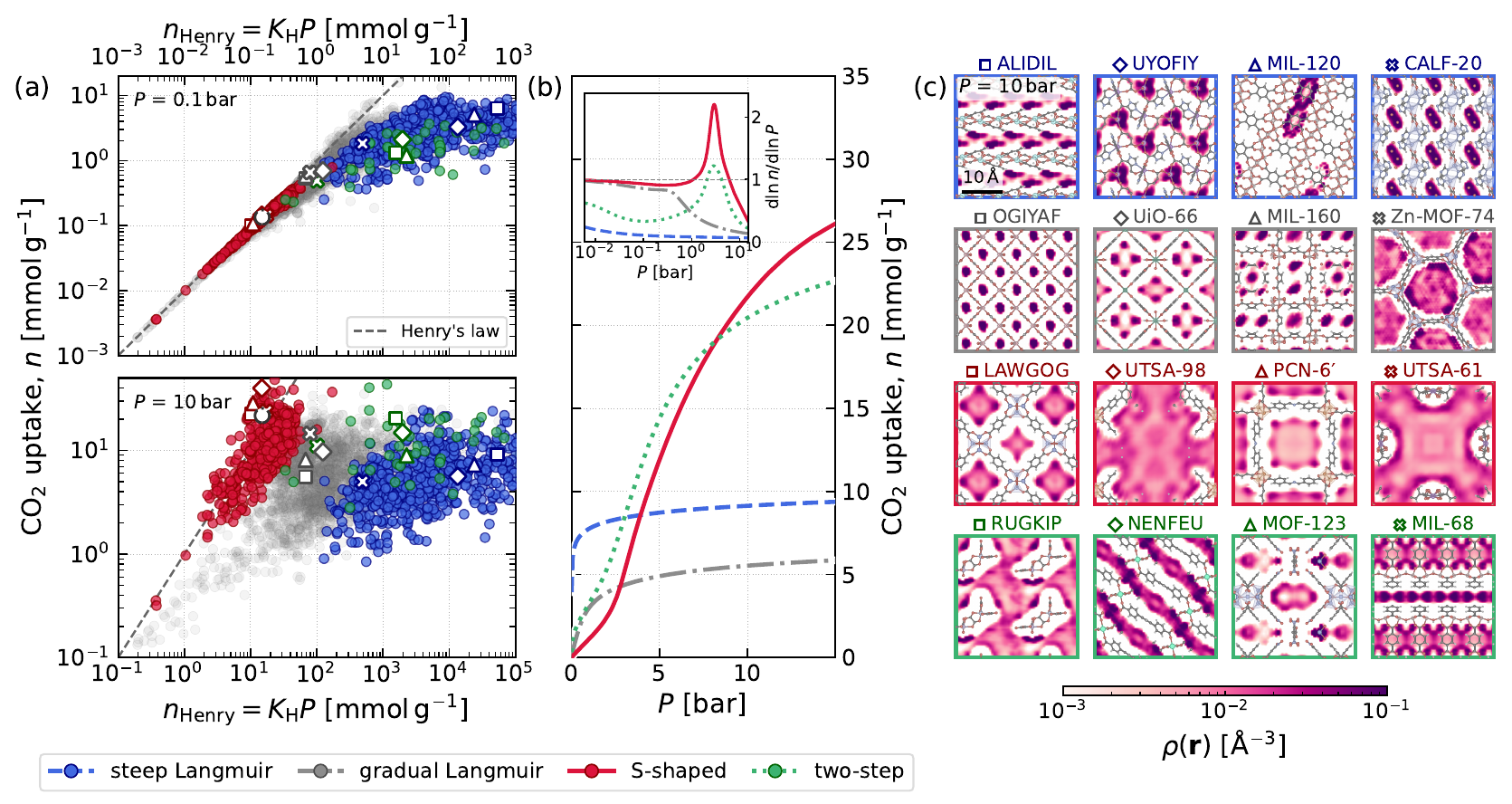}
  \caption{\tbf{Isotherm classes across the CoRE MOF 2025 database.} At 298\,K, $\mathrm{d}\ln n/\mathrm{d}\ln P$ distinguishes four cDFT isotherm classes (UMA-m, PBE-D3 CO$_2$). At 10\,bar (bottom), uptake deviates more strongly from Henry's law ($n=K_\mrm{H}P$, dashed) than at 0.1\,bar (top) in (a). The representative curves in (b) range from steep and gradual Langmuir isotherms to S-shaped and two-step profiles, with their slopes in the inset. For the MOFs marked in white in (a), the 10\,bar CO$_2$ density maps in (c) span four examples per class (rows), each sliced through the deepest $U_\mrm{ext}$ site or a metal node.}
  \label{fig:MOF_screen}
\end{figure*}

Following the validation of our framework in the previous section, we now turn to the adsorption thermodynamics for MOFs where these properties have yet to be characterized. This lets us classify a wide range of MOFs by their adsorption and provide a database for the MOF community. The 5,591 MOFs we screened are those in the CoRE MOF 2025 database \cite{Zhao2025CoREMOFDB} that have been experimentally synthesized. In this section, we show the results for UMA-m with the PBE-D3 CO$_2$ functional, but we also provide the full isotherms with uMOF-MH, MACE-DAC and UFF.

In Figure~\ref{fig:MOF_screen}(a) we show the computed CO$_2$ uptake, $n$, of the CoRE MOFs at 298\,K against the uptake from Henry's law, $n_\mrm{Henry} = K_\mrm{H}P$. While the Henry constants $K_\mrm{H}$ are widely used to rank MOFs, they characterize only the dilute limit where uptake varies linearly with pressure. Starting at a lower pressure, $P = 0.1\,$bar, in the top panel of Fig.~\ref{fig:MOF_screen}(a) we observe that the MOFs with lower uptake values, $n<10$\,mmol\,g$^{-1}$, lie close to Henry's law such that $K_\mrm{H}$ alone predicts the uptake. However, at 10\,bar, this is no longer the case: MOFs with the same $K_\mrm{H}$ can differ in uptake by nearly two orders of magnitude, since $K_\mrm{H}$ does not encode how the pores fill beyond the dilute limit. A screening of MOFs based on the Henry constant alone therefore cannot distinguish between different adsorption mechanisms.

Fig.~\ref{fig:MOF_screen}(b) shows the isotherms of four representative MOFs, whose shapes differ long before they saturate. To classify the MOFs of the database by shape, we use the slope of the room-temperature isotherm, $\mathrm{d}\ln n/\mathrm{d}\ln P$ (inset). This slope equals 1 in the dilute limit, falls toward 0 as sites or pores fill and rises above 1 where CO$_2$--CO$_2$ attraction makes the uptake grow faster than the pressure.

Each MOF is assigned to one of four classes by three tests in turn. A MOF is ``two-step'' when its slope drops and then rises again. Otherwise, it is ``S-shaped'' when its slope clearly exceeds 1 at some pressure of 1\,bar or above, reminiscent of the step-shaped isotherms sought for small-swing capture \cite{mcdonald2015cooperative, danaci2023evaluating}. The remaining isotherms are Langmuir-like, concave over the whole range, i.e., $\mathrm{d}\ln n/\mathrm{d}\ln P < 1 $, and can be further classified through a fit to the dual-site Langmuir model. A MOF is ``steep Langmuir'' when the fit needs two well-separated components: a steep one that half-fills at a pressure more than 50 times lower than the other and holds more than 20\,\% of the capacity. Otherwise, it is ``gradual Langmuir''. Most MOFs are gradual Langmuir. S-shaped and steep Langmuir MOFs each make up a smaller share, and two-step MOFs are rare.

We can also use the CO$_2$ density in the MOFs obtained from our cDFT approach to shed light on the microscopic origin of each isotherm class. Figure~\ref{fig:MOF_screen}(c) shows the CO$_2$ density for representative MOFs from the database at 10\,bar and Fig.~S12 the size and strength of the strongest binding site across the database. Steep Langmuir MOFs hold strong single binding sites, typically exposed metals, in narrow pores, with a pore diameter of $\sim$4\,\AA{} at the strongest site, such that CO$_2$ packs into small pockets. In contrast, S-shaped MOFs have wider pores, $\sim$7\,\AA, with weakly binding walls and no strong single sites. They are rich in bridged metal clusters (M--X--M, Zn$_4$O), as in the IRMOF-type Zn$_4$O-pcu frameworks and Cu paddlewheels, with few polar linkers, and CO$_2$ fills the pore evenly. Two-step MOFs combine large pores with strong sites, which fill first: exposed metal sites before the pore, or small pockets before the large one (e.g., MIL-68 \cite{volkringer2008mil68}). Gradual Langmuir MOFs lie in between, with small to medium pores.

\section{Working capacities across swing adsorption processes}

\begin{figure*}[t]
  \includegraphics[width=\linewidth]{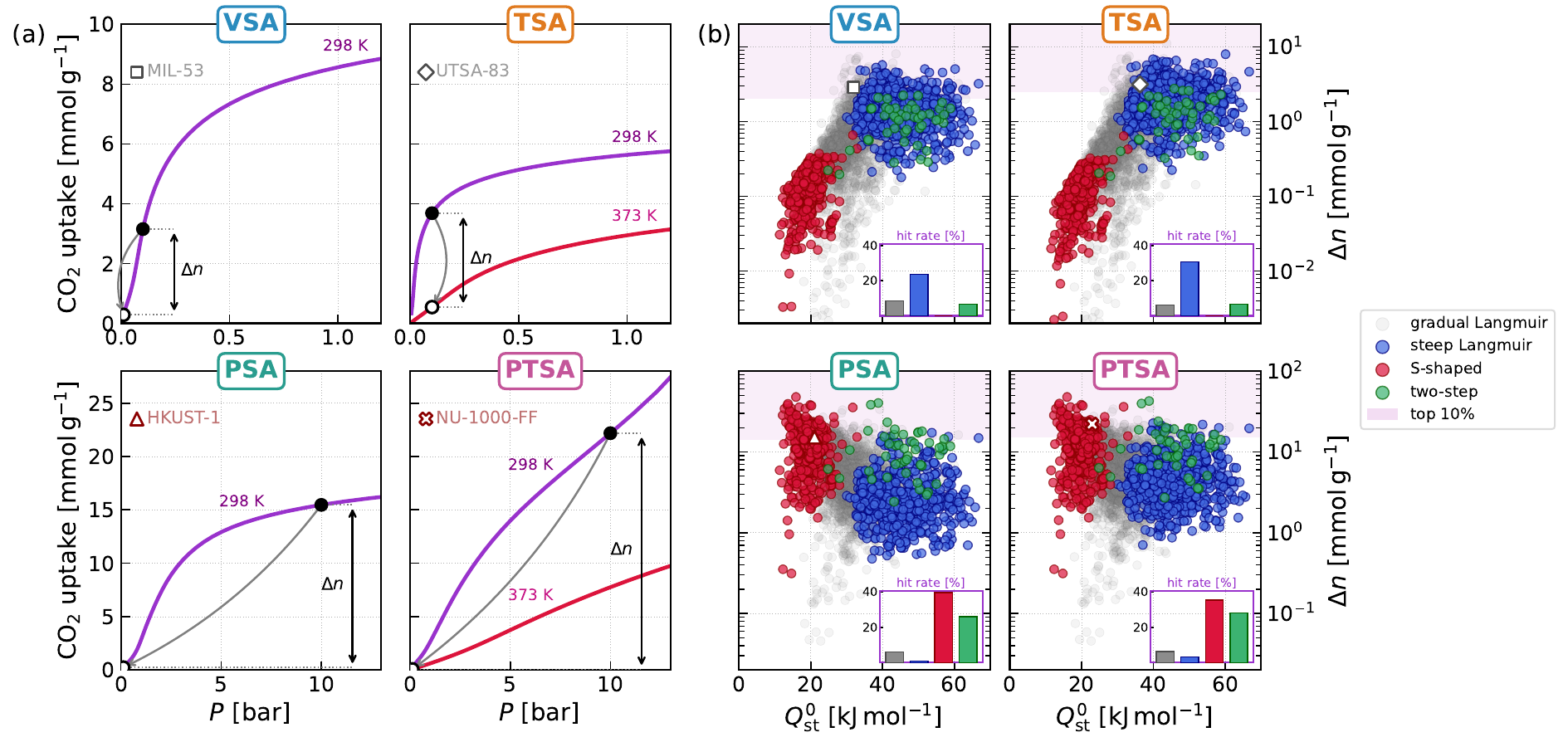}
  \caption{\tbf{CO$_2$ working capacities for vacuum (VSA), temperature (TSA), pressure (PSA) and pressure--temperature (PTSA) swing adsorption.} The UMA-m isotherms in (a) illustrate the working capacity $\Delta n$, the uptake difference between adsorption (filled symbols) and desorption (open symbols). At 298\,K, VSA lowers the pressure from 0.1 to 0.01\,bar and PSA from 10 to 0.1\,bar. TSA raises the temperature from 298 to 373\,K at 0.1\,bar, while PTSA combines this heating with the PSA pressure drop. Across all MOFs, (b) relates $\Delta n$ to $Q^0_\mrm{st}$ by isotherm class (Fig.~\ref{fig:MOF_screen}). The fraction of each class reaching the shaded top 10\,\% (insets) is highest for steep Langmuir isotherms under VSA and TSA and for S-shaped isotherms under PSA and PTSA.}
  \label{fig:swing_capacity}
\end{figure*}

Designing a material for CO$_2$ capture requires tailoring it for a desired swing process under the required operating conditions: CO$_2$ is adsorbed at one thermodynamic condition and released at another, which regenerates the material. Current regeneration technologies rely on pressure swing adsorption (PSA), used mainly for pre-combustion CO$_2$ separation~\cite{nandi2015singleligand, Herm2011PrecombustionCapture}, and on vacuum (VSA) and temperature swing adsorption (TSA), used for post-combustion capture~\cite{Trickett2017MOFCO2, Mason2011PostcombustionTSA}, together with hybrid cycles such as pressure--vacuum (PVSA) \cite{maring2013new, maring2014corrigendum, danaci2020exploring} and temperature--vacuum swing adsorption (TVSA), the latter widely used for direct air capture \cite{wurzbacher2012concurrent, shi2023temperaturevacuum}. In PSA and VSA, CO$_2$ is adsorbed at a given pressure and released by lowering it. In TSA, it is instead released by heating the adsorbent, and the performance depends on both temperatures and on the CO$_2$ partial pressure \cite{raganati2020capture}. A pressure--temperature swing (PTSA) does both at once. In all cases, the figure of merit is the working capacity, i.e., the uptake at the adsorption condition minus that at the desorption condition \cite{mahajan2022recent}, $\Delta n = n(P_\mrm{ads}, T_\mrm{ads}) - n(P_\mrm{des}, T_\mrm{des})$.

\begin{figure}[t]
  \includegraphics[width=\linewidth]{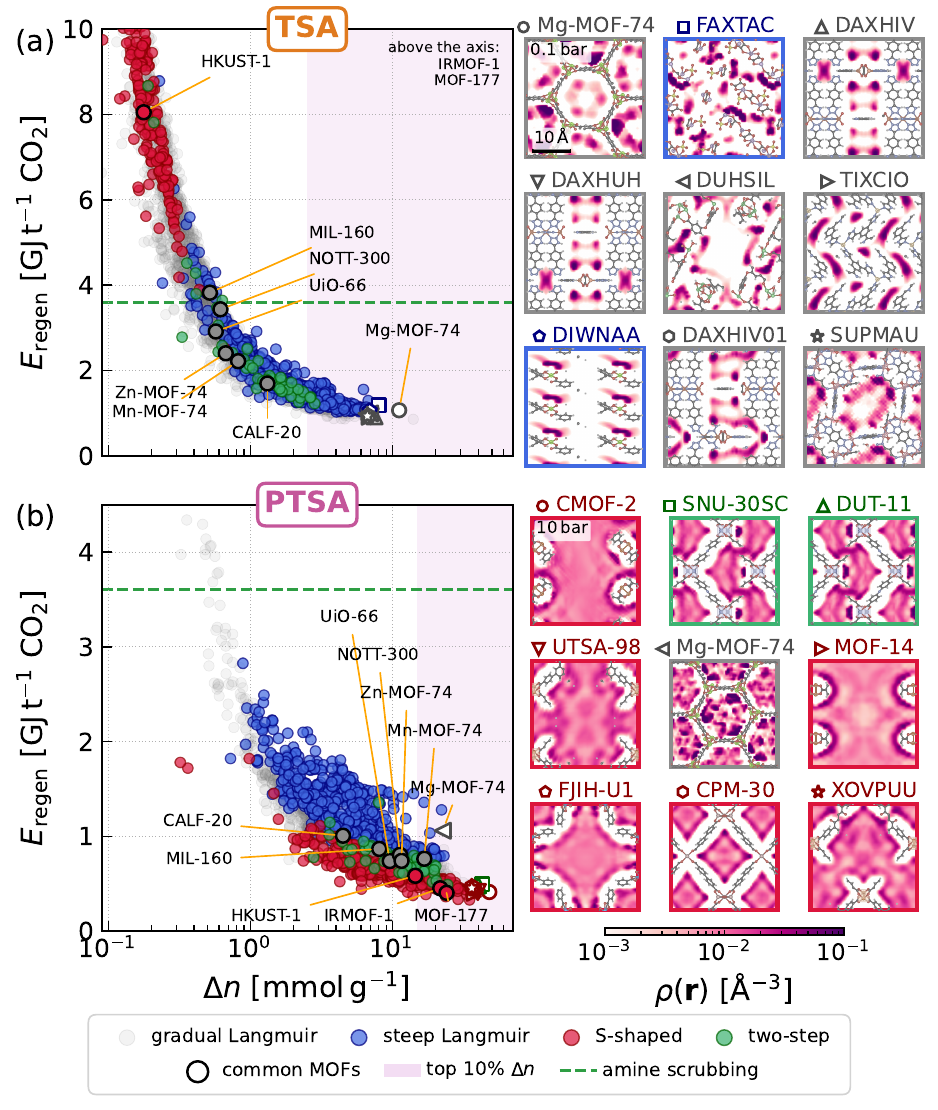}
  \caption{\tbf{Regeneration energy for TSA and PTSA.} Regeneration energy $E_\mrm{regen}$ (Eq.~\eqref{eq:eregen}) generally decreases as working capacity $\Delta n$ increases for both TSA (a) and PTSA (b). For both swings, all MOFs in the top 10\,\% by $\Delta n$ have regeneration energies below the amine scrubbing benchmark of 3.6\,GJ\,t$^{-1}$ CO$_2$ \cite{rochelle2009amine}. The density maps give the CO$_2$ distribution at adsorption conditions for representative high-capacity MOFs with regeneration energies below the benchmark. Mg-MOF-74, the top TSA candidate, is included for both swings.}
  \label{fig:regen}
\end{figure}

Here, we consider four swings, VSA, TSA, PSA and PTSA, which are shown for representative materials from the database in Fig.~\ref{fig:swing_capacity}(a). All of the swings undergo adsorption at room temperature and the heated swings desorb at 373\,K, within the 363--413\,K range commonly used in MOF screening \cite{boyd2019datadriven, Mason2011PostcombustionTSA}. Since our functional was trained up to 500\,K, which covers most CO$_2$ capture conditions \cite{bae2011development,wilmer2012structureproperty}, the working capacity of any such swing follows directly for every MOF in the database.

In Fig.~\ref{fig:swing_capacity}(b), we show the working capacity of every MOF for each swing against the isosteric heat of adsorption $Q^0_\mrm{st}$. We immediately see that the isotherm classes of Sec.~\ref{sec:classification} separate clearly. For VSA and TSA, the MOFs with the highest working capacities ($\Delta n$) are steep Langmuir MOFs with strong binding sites, whose isotherms rise most steeply at low pressure, whereas S-shaped MOFs perform poorly. For PSA and PTSA, where CO$_2$ is adsorbed at high pressure, the order reverses. A high working capacity then requires uptake that keeps rising at high pressure, so a gradual rise that keeps filling the pore pays off: S-shaped and two-step MOFs perform best, whereas steep Langmuir MOFs, whose strong sites remain occupied at the desorption step, fall behind. Notably, gradual Langmuir MOFs, which make up the majority of the database, are rarely top performers in any swing.  Put simply, the class of MOF required to achieve the highest working capacity depends on the swing. Small pores with strong binding sites, often open metal sites, are best for low-pressure capture, whereas large pores with weakly binding walls, alone or with strong sites as in two-step MOFs, pay off in high-pressure swings.

It is important to stress that the working capacity is only a first screening metric, as simple metrics rank MOFs less reliably than optimizing the full process \cite{rajagopalan2016adsorbent, farmahini2021performancebased}, and process-based metrics \cite{khurana2016adsorbent, leperi2019development} or models trained on process optimizations \cite{burns2020prediction} perform better. Importantly, because our cDFT approach can efficiently compute isotherms at any temperature and pressure, they can be fed directly into such process models, as recently demonstrated for CH$_4$/N$_2$ separation \cite{Granderath2026ProcessScreening}. This paves the way to screen entire databases at the process level \cite{khurana2017integrated}.

\section{Regeneration energy and promising MOFs}

A large working capacity alone is not sufficient, since optimizing a swing process necessitates that the material can release the captured CO$_2$ with little energy. In Fig.~\ref{fig:regen} for the thermal swings from 298 to 373\,K, we show the regeneration energy per tonne of CO$_2$ 
calculated from~\cite{Mason2011PostcombustionTSA, lin2012insilico} 
  \begin{equation}
    E_\mrm{regen} = Q_\mrm{des} + \frac{Q_\mrm{sens}}{\Delta n}.
    \label{eq:eregen}
  \end{equation}
The first term in Eq.~\eqref{eq:eregen} is the heat to desorb the CO$_2$ that the swing releases, $Q_\mrm{des} = (\Delta n)^{-1}\int_{n_\mrm{des}}^{n_\mrm{ads}}\!\mrm{d}n\, Q_\mrm{st}(n)$, where $Q_\mrm{st}(n)$ is computed from Eq.~\eqref{eq:qst}. The second is the sensible heat per gram of framework, $Q_\mrm{sens} = \int_{T_\mrm{ads}}^{T_\mrm{des}}\!\mrm{d}T\, c_p(T)$, where the heat capacity $c_p(T)$ for each MOF was taken from the CoRE MOF 2025 database \cite{Zhao2025CoREMOFDB, Moosavi2022HeatCapacity}. 

Figure~\ref{fig:regen} shows the regeneration energy and the working capacity of materials in the CoRE database for TSA (Fig.~\ref{fig:regen}(a)) and PTSA (Fig.~\ref{fig:regen}(b)) processes with each MOF colored by its class. The green line marks amine scrubbing \cite{rochelle2009amine}, and MOFs below it regenerate more cheaply than the industrial benchmark. The two terms of Eq.~\eqref{eq:eregen} control $E_\mrm{regen}$ differently in the two swings. In TSA the working capacities are small, so the sensible heat, which scales as $1/\Delta n$, dominates such that $E_\mrm{regen}$ falls steeply with $\Delta n$. In contrast, PTSA releases CO$_2$ both by heating and by lowering the pressure. Its working capacities are consequently an order of magnitude larger, the sensible heat becomes small and $E_\mrm{regen}$ instead follows the heat of desorption $Q_\mrm{des}$. The two swings favor different classes of MOFs. For TSA only about half of all MOFs regenerate below amine scrubbing and the best combinations of low regeneration energy and high uptake come mostly from steep and gradual Langmuir MOFs. In contrast, for PTSA the classes of MOFs that lead to optimal performance shift to those classified as S-shaped MOFs and two-step MOFs. In these MOFs the released CO$_2$ is weakly bound and their strong sites remain occupied while the pore empties. In PTSA, nearly every MOF regenerates below the amine benchmark, although this comparison neglects the work of compressing the feed to 10\,bar, and the benchmark refers to post-combustion capture from flue gas.

The most established MOFs from our experimental comparison that have been suggested for TSA and PTSA processes are labeled in Fig.~\ref{fig:regen}.  For TSA, the MOF-74 analogs, with their open metal sites, regenerate far more cheaply than the large-pore MOF-177 and IRMOF-1, consistent with the comparison of Mg-MOF-74 and MOF-177 by Mason et al.\ \cite{Mason2011PostcombustionTSA}, while CALF-20, which has recently been scaled up as an industrial physisorbent~\cite{lin2021scalable}, has better performance in both $E_\mrm{regen}$ and $\Delta n$. For PTSA, the order reverses, with the large-pore, S-shaped MOFs of the set, e.g., IRMOF-1, MOF-177 and HKUST-1, going from the highest regeneration cost in TSA to the lowest in PTSA, in line with the large high-pressure capacity that made MOF-177 a candidate for pre-combustion capture \cite{Herm2011PrecombustionCapture}.

We now consider the top-performing MOFs, i.e., those with the highest uptake and lowest regeneration energy, predicted for each process obtained from our calculations on the CoRE database. The CO$_2$ densities for these materials are shown in Fig.~\ref{fig:regen}(c) for TSA (top) and PTSA (bottom) swings with the box color denoting their classification. The top-20 for each swing, with our calculated $K_\mrm{H}$, $Q^0_\mrm{st}$ and their isotherm class, are listed in Tables~S2 and S3. For TSA, the top-20 MOFs concentrate CO$_2$ in small pockets, with the well-established Mg-MOF-74 \cite{caskey2008mof74} ranked first and analogs of the HKUST-1, NNU-21, HPU-20 and CHD-1 families also among them. The PTSA candidates are instead mostly large-pore S-shaped and two-step MOFs that fill their whole pore, alongside a few strong-site frameworks with a large capacity. Owing to the comparatively high computational cost of the UMA-m MLIP, we also performed the screening with the more efficient uMOF-MH MLIP (Sec.~S10), which keeps nearly all of the UMA-m candidates within the best 10\,\% of MOFs.

Our ability to efficiently identify MOFs with desired uptake and regeneration energies, and the fact that highly optimized CO$_2$ capture materials such as Mg-MOF-74 emerge naturally as top candidates, are promising. However, we stress that these rankings rest on our particular choice of performance metric based purely on uptake and regeneration energy for pure CO$_2$ isotherms of rigid MOFs. Competition with N$_2$ and water, which can reorder candidates for wet flue gas \cite{boyd2019datadriven}, framework flexibility and the kinetics and heat effects within an adsorption column are not included in our current approach. Our pure CO$_2$ results nonetheless apply most directly where CO$_2$ is nearly the only adsorbing gas, e.g., pre-combustion capture from weakly adsorbing H$_2$ at 5--40\,bar~\cite{Herm2011PrecombustionCapture}, the second stage of multistage capture~\cite{Ling2016VSA} and CO$_2$--MOF heat pumps, which cycle pure CO$_2$ at 8--30\,bar and favor S-shaped isotherms~\cite{Shamim2022HybridHeatPump, Hu2026MOFHeatExchanger}.

\section{Conclusions and outlook}

We have presented a framework that combines a neural classical density functional of the CO$_2$ fluid, learned from molecular dynamics simulations, with external potentials of the CO$_2$--MOF interactions from foundational MLIPs. Our cDFT approach yields the CO$_2$ adsorption isotherm of a MOF across the entire range of loadings, from the Henry regime to pore filling, at the accuracy of the underlying MLIP---currently the upper rungs of the DFT hierarchy. We demonstrated that cDFT closely reproduces isotherms from GCMC using the UFF classical force field, while avoiding the need to perform grand canonical simulations. Benchmarking seven foundational MLIPs and the UFF force field against the experimental isotherms of 20 well-characterized MOFs, we found that they differ widely in how well they capture adsorption. UMA-m is the most accurate, reproducing the measured isotherms to within a factor of $\sim$1.5, and uMOF-MH achieves similar accuracy at an order of magnitude lower cost, making it an attractive choice when compute is limited. In contrast, UFF misses strong binding sites, whereas MACE-DAC, fine-tuned for direct air capture, overbinds them.

Applying our framework to over 5,000 MOFs from the CoRE MOF 2025 database, we used their full isotherms to reveal what the Henry constant alone cannot. We classified the MOFs into four isotherm shapes, set by their pore size and binding motifs, which rather than the binding strength alone determine which MOFs suit a given swing. Low-pressure swings favor steep Langmuir MOFs, whose small pores hold strong binding sites, whereas high-pressure swings favor S-shaped MOFs, whose large pores with weakly binding walls fill cooperatively. Two-step MOFs, which combine both, perform well in high-pressure swings, while the gradual Langmuir MOFs that make up most of the database are under-represented throughout. Which class has the lowest regeneration energy also depends on the swing process. In TSA, MOFs with strong binding sites, such as Mg-MOF-74, are the most efficient, whereas in PTSA, where the regeneration energy is dominated by the heat of desorption, S-shaped and two-step MOFs, such as UTSA-98 and DUT-11, which release weakly bound CO$_2$, are. Pore size and the presence of strong binding sites thus emerge as the structural handles for tailoring a MOF to its process. 

An appealing feature of our approach is that the functional describes the fluid independently of the MOF framework. Future improvements in foundational MLIPs can therefore be directly translated into improved isotherm, isosteric heat and regeneration energy predictions that can be fed into process models \cite{Granderath2026ProcessScreening}, turning a screen of working capacities into one of process performance. Looking forward, we envisage several natural extensions. Functionals that resolve molecular orientations explicitly \cite{Yang2025OperatorLearning} could go beyond the orientation-averaged external potential, and multicomponent functionals could describe mixtures of CO$_2$ with N$_2$ and water \cite{thiele2026efficient}. This approach could also be extended to flexible MOF frameworks, to direct air capture, where the low partial pressure puts the strongest sites to the test \cite{sanzperez2016direct, deutz2021lifecycle}, and to capture at high temperature \cite{rohde2024hightemperature}. Using the same learned density functional in dynamical DFT could give access to breakthrough and other nonequilibrium adsorption phenomena \cite{ram2025learned}. Finally, the ability to generate adsorption thermodynamics with ab initio accuracy efficiently under a wide range of conditions opens the door to addressing the inverse question of which isotherm a given swing requires and which frameworks deliver it \cite{khurana2017integrated}, potentially heralding a new generation of design for carbon capture materials.

\section*{Data Availability}

Data and code supporting the findings of this study will
be openly available upon publication of the manuscript.

\section*{Acknowledgments}

A.T.B. is supported by a Stanford Science Fellowship. This work used the Delta system at the National Center for Supercomputing Applications through allocation CHEM260198 from the Advanced Cyberinfrastructure Coordination Ecosystem: Services \& Support (ACCESS) program, which is supported by U.S. National Science Foundation grants \#2138259, \#2138286, \#2138307, \#2137603 and \#2138296. Some of the computing for this project was performed on the Sherlock cluster. We would like to thank Stanford University and Stanford Research Computing for providing computational resources and support that contributed to these research results.

\bibliography{references}

\end{document}


\title{Supporting information: Characterizing adsorption thermodynamics in carbon capture materials: an efficient first-principles approach}

\author{Anna T. Bui}
\email{annatbui@stanford.edu}
\affiliation{Department of Chemistry, Stanford University, Stanford, California, 94305, USA}

\author{Grant M. Rotskoff}
\email{rotskoff@stanford.edu}
\affiliation{Department of Chemistry, Stanford University, Stanford, California, 94305, USA}

\author{Thomas E. Markland}
\email{tmarkland@stanford.edu}
\affiliation{Department of Chemistry, Stanford University, Stanford, California, 94305, USA}

\date{\today}

\maketitle

\tableofcontents


\section{Theoretical framework}

We describe the CO$_2$ fluid by the number density $\rho(\mbf{r})$ of the carbon site of each molecule. In the grand canonical ensemble, the grand potential functional is
\begin{equation}
  \varOmega_U([\rho];T) = \mcal{F}_{\mrm{id}}([\rho];T) + \mcal{F}_{\mrm{ex}}([\rho];T) + \int\!\mrm{d}\mbf{r}\,\rho(\mbf{r})\left[U_\mrm{ext}(\mbf{r})-\mu\right],
  \label{eq:omega}
\end{equation}
where the ideal intrinsic Helmholtz free energy is $\mcal{F}_{\mrm{id}}([\rho];T) = k_\mrm{B}T\int\mrm{d}\mbf{r}\,\rho(\mbf{r})\left[\ln\left(\zeta^{-1}\Lambda^3\rho(\mbf{r})\right)-1\right]$, with $\Lambda$ the thermal de Broglie wavelength and $\zeta$ an intramolecular partition function accounting for the orientational and internal degrees of freedom of each molecule, $U_\mrm{ext}(\mbf{r})$ the external potential acting on the carbon site, $\mu$ the chemical potential and $T$ the temperature. The excess intrinsic free energy $\mcal{F}_{\mrm{ex}}$ generates the one-body direct correlation functional,
\begin{equation}
  c^{(1)}(\mbf{r};[\rho],T) = -\frac{\delta \beta \mcal{F}_{\mrm{ex}}([\rho];T)}{\delta\rho(\mbf{r})},
\end{equation}
and minimizing Eq.~\ref{eq:omega} gives the Euler--Lagrange equation for the equilibrium density $\rho_\mrm{eq}(\mbf{r})$,
\begin{equation}
  \rho_\mrm{eq}(\mbf{r}) = \frac{\zeta}{\Lambda^{3}}\exp\left[-\beta\left(U_\mrm{ext}(\mbf{r})-\mu\right) + c^{(1)}(\mbf{r};[\rho_\mrm{eq}],T)\right].
  \label{eq:EL}
\end{equation}
In what follows, we drop the subscript and write $\rho$ for the equilibrium density. When rearranged,
\begin{equation}
  c^{(1)}(\mbf{r};[\rho],T) = \ln\!\left(\zeta^{-1}\Lambda^3\rho(\mbf{r})\right) + \beta U_\mrm{ext}(\mbf{r}) - \beta\mu,
  \label{eq:c1-learn}
\end{equation}
the Euler--Lagrange equation provides the training target from simulation (Sec.~\ref{sec:training}). We set $\zeta^{-1}\Lambda^3 = 1$\,\AA$^3$ throughout, which only shifts $\mu$.

The bulk equation of state follows from the same functional. For the uniform fluid of density $\rho_\mrm{b}$, the chemical potential is $\beta\mu_\mrm{b} = \ln(\zeta^{-1}\Lambda^3\rho_\mrm{b}) - c^{(1)}([\rho_\mrm{b}],T)$, and the pressure, $P = -\varOmega_U/V$ at $U_\mrm{ext}=0$ with $V$ the volume, is
\begin{equation}
  \beta P = \rho_\mrm{b}\left[1 - c^{(1)}([\rho_\mrm{b}],T)\right] - \frac{\beta\mcal{F}_\mrm{ex}([\rho_\mrm{b}];T)}{V}.
  \label{eq:pressure}
\end{equation}
The excess free energy is obtained from $c^{(1)}$ by functional line integration,
\begin{equation}
  \beta\mcal{F}_\mrm{ex}([\rho];T) = -\int_0^1\!\mrm{d}\alpha\int\!\mrm{d}\mbf{r}\,\rho(\mbf{r})\,c^{(1)}(\mbf{r};[\alpha\rho],T),
  \label{eq:fli}
\end{equation}
which, for the uniform fluid, reduces to $\beta\mcal{F}_\mrm{ex}([\rho_\mrm{b}];T)/V = -\rho_\mrm{b}\int_0^1\mrm{d}\alpha\,c^{(1)}([\alpha\rho_\mrm{b}],T)$.

\section{Neural cDFT training}
\label{sec:training}

\textit{Training data.} We performed canonical MD simulations of CO$_2$ in randomly generated 3D external potentials with LAMMPS \cite{Thompson2022LAMMPS}, in a cubic periodic box of length 33\,\AA. The external potential acts on the carbon site and is a sum of up to four cosines per Cartesian direction,
\begin{equation}
  V_\mrm{ext}(\mbf{r}) = \sum_{a\in\{x,y,z\}}\sum_{n} A_{a,n}\cos\!\left(\frac{2\pi k_{a,n} r_a}{L}+\phi_{a,n}\right) + \sum_{m} D_m \prod_{a} g\!\left(\frac{h_{m,a}-|r_a-c_{m,a}|}{e_m}\right),
\end{equation}
with $g(u) = [1+\exp(-2u)]^{-1}$, plus up to eight smooth rectangular wells and barriers. Amplitudes $A_{a,n}$ and depths $D_m$ are drawn uniformly from $[-4.2,4.2]$ and $[-16.7,16.7]$\,kJ\,mol$^{-1}$, wave numbers $k_{a,n}\in\{1,\dots,4\}$, phases from $[0,2\pi)$, well half-widths $h_{m,a}$ from $[0.5,5]$\,\AA{} and edge widths $e_m$ from $[0.05,1]$\,\AA. Each simulation has $N\in[70,1024]$ molecules and a temperature drawn uniformly from 220 to 500\,K. For TraPPE CO$_2$ \cite{Potoff2001VaporLiquid}, molecules are rigid, Lennard-Jones interactions are cut and shifted at 15\,\AA, electrostatics are computed with PPPM to an accuracy of $10^{-5}$, and the temperature is maintained with a Nos\'e--Hoover thermostat with a damping time of 200\,fs. Each simulation was equilibrated for 25\,ps with a time step of 0.5\,fs, followed by production with a time step of 1\,fs for 4 to 130\,ns, typically tens of ns. The carbon site density was sampled on a grid of spacing 0.5\,\AA{} every 1\,ps. We discarded simulations in which the fluid had frozen, identified by a root-mean-square displacement of the carbon atoms below 10\,\AA{} over 1\,ns, against several tens of \AA{} in the liquid. We also performed bulk simulations of 10\,ns each at six temperatures from 250 to 500\,K and 36 densities each, whose pressures enter the loss below.

For PBE-D3, the first 128 simulations were repeated with the same external potentials, $N$ and $T$, with a MACE potential from Ref.~\onlinecite{Bui2026MultiscaleLiquids} trained with data from Ref.~\onlinecite{Mathur2023CO2PhaseBehavior}. The CO$_2$ molecules are  flexible so a time step of 0.5\,fs was used. Each inhomogeneous simulation has approximately 15\,ns of sampling in total. Bulk PBE-D3 pressures come from 1.5\,ns simulations.

\textit{Network and loss.} The network takes as input the density in a sphere of radius $r_\mrm{c} = 6$\,\AA{} around each voxel, i.e., 7153 voxels at 0.5\,\AA, together with $T$, and returns $c^{(1)}$ at that voxel. It is a fully connected network implemented in PyTorch \cite{Paszke2017PyTorch}, with hidden layers of 128, 64 and 32 nodes and softplus activations, approximately $9\times10^5$ parameters in total. Since $\mu$ of a canonical simulation is not known, we assign one trainable scalar $\mu$ to each training simulation \cite{Sammuller2026ChemicalPotential, Bui2026MultiscaleLiquids}, and minimize the loss
\begin{equation}
\begin{aligned}
  \mcal{L}(\theta,\{\mu_s\}) = {} & \left\| c^{(1)}_\theta(\mbf{r};[\rho],T) - \ln\!\left(\zeta^{-1}\Lambda^3\rho(\mbf{r})\right) - \beta U_\mrm{ext}(\mbf{r}) + \beta\mu_s \right\|^2 \\
  & + \lambda_P \left\| \frac{P_\theta - P_\mrm{MD}}{|P_\mrm{MD}| + P_0} \right\|^2
  + \lambda_0 \left\| c^{(1)}_\theta([\rho=0],T) \right\|^2
  + \lambda_\nabla \left\| \nabla_{\rho} c^{(1)}_\theta \right\|^2 ,
\end{aligned}
  \label{eq:loss}
\end{equation}
where $\|\cdot\|^2$ denotes the mean square over occupied voxels in the first and last terms, over bulk state points in the second and over temperatures in the third.
The first term is the squared residual of Eq.~\ref{eq:c1-learn} over the training simulations, with $c^{(1)}_\theta$ the output of the network with parameters $\theta$ and $\mu_s$ the latent chemical potential of simulation $s$, both optimized together. The second is the relative error of the bulk pressure of Eq.~\ref{eq:pressure}, $P_\theta$, against that of the bulk simulations, $P_\mrm{MD}$, with $P_0 = 10$\,atm and $\lambda_P = 1$. The third, with $\lambda_0 = 1$, fixes the gauge $c^{(1)}([\rho=0],T)=0$ at 16 temperatures, and the last is a small penalty, $\lambda_\nabla = 10^{-6}$, on the gradient of the output with respect to the input density. Voxels with $\rho > 0.05$\,\AA$^{-3}$ are oversampled to make up 10\,\% of each epoch.

\textit{Training.} The TraPPE functional was trained on 100 simulations, with 41 for validation, for 50 epochs of $5\times10^5$ voxels in batches of 1024, with the Adam optimizer \cite{Kingma2014Adam} and a learning rate decreasing by 5\,\% per epoch from $10^{-3}$. The best validation loss, 0.070, is reached at the last epoch, and the bulk pressure is reproduced to within 40--90\,atm from 250 to 450\,K. The PBE-D3 functional was fine-tuned from it on 76 simulations, with 25 for validation: only the layers after the first are trained, with a learning rate of $10^{-4}$, and $10^{-3}$ for $\mu$, and the pressure term of Eq.~\ref{eq:loss}, with $\lambda_P = 0.5$, against the bulk equation of state of the planar PBE-D3 functional of Ref.~\onlinecite{Bui2026MultiscaleLiquids}. Fine-tuning lowers the validation loss from 0.30, that of the TraPPE functional, to 0.15. Each training takes under 30\,min on the GPU of a single Apple M5 chip.

\textit{Using the functional.} Once trained, the functional is used to solve Eq.~\ref{eq:EL} on a periodic grid of spacing 0.5\,\AA, the resolution of the functional. The network is evaluated on the whole grid at once, with its first layer computed as a correlation by fast Fourier transform. We solve Eq.~\ref{eq:EL} for $\ln(\zeta^{-1}\Lambda^3\rho)$ by Picard iteration with a mixing parameter of 0.01, accelerated with Anderson mixing \cite{Anderson1965Iterative} over the last five iterations, which is equivalent to the direct inversion in the iterative subspace (DIIS) of electronic-structure calculations \cite{Pulay1980DIIS}, until $\max|\Delta\rho| < 10^{-6}$\,\AA$^{-3}$. For a sequence of chemical potentials, each solve is started from the converged density at the previous $\beta\mu$, with adaptive steps of at most 1 in $\beta\mu$. To prevent the iteration from diverging in the deepest wells at high pressure, where the density would otherwise reach values far outside the training data, we bound the local density softly from above by adding $\ln(1-\rho/\rho_\mrm{cap})$ to the exponent of Eq.~\ref{eq:EL}, with $\rho_\mrm{cap}=0.2$\,\AA$^{-3}$. This changes $c^{(1)}$ by less than 0.1 at the density of the bulk liquid.

\section{Neural cDFT validation}
\label{sec:validation}

\begin{figure*}[h]
  \includegraphics[width=\linewidth]{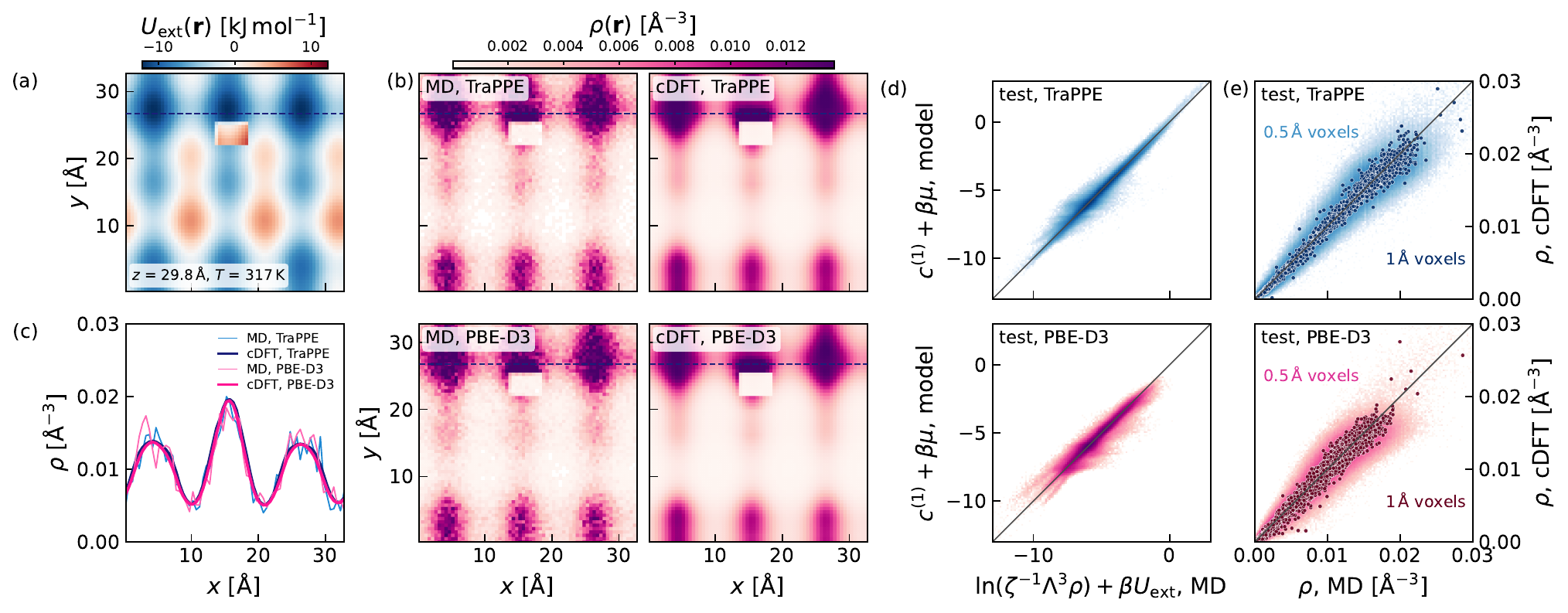}
  \caption{\tbf{Held-out test set: MD against cDFT for both CO$_2$ functionals.} One simulation of confined CO$_2$ left out of the training of both functionals, for the TraPPE (top row) and PBE-D3 (bottom row) CO$_2$ model, at the same external potential, temperature and number of molecules. (a) The external potential $U_\mrm{ext}$ in one slice. (b) The CO$_2$ density in that slice from MD and from a single cDFT solve at the simulation's own chemical potential and temperature. (c) A 1D cut through it, along the regular wells. (d) Parity of $c^{(1)} + \beta\mu$ over the held-out test set: $\ln(\zeta^{-1}\Lambda^3\rho) + \beta U_\mrm{ext}$ from MD against the trained functional, omitting voxels with $\rho_\mrm{MD} < 10^{-4}\,$\AA$^{-3}$. (e) Parity of the density over the held-out test set, cDFT solved at that $\beta\mu$ against MD, on the 0.5\,\AA{} grid of the simulations and with both densities averaged to 1\,\AA{} voxels (dark points, a random subsample), which removes most of the MD sampling noise.}
\label{figS1}
\end{figure*}

\begin{figure*}[h]
  \includegraphics[width=0.95\linewidth]{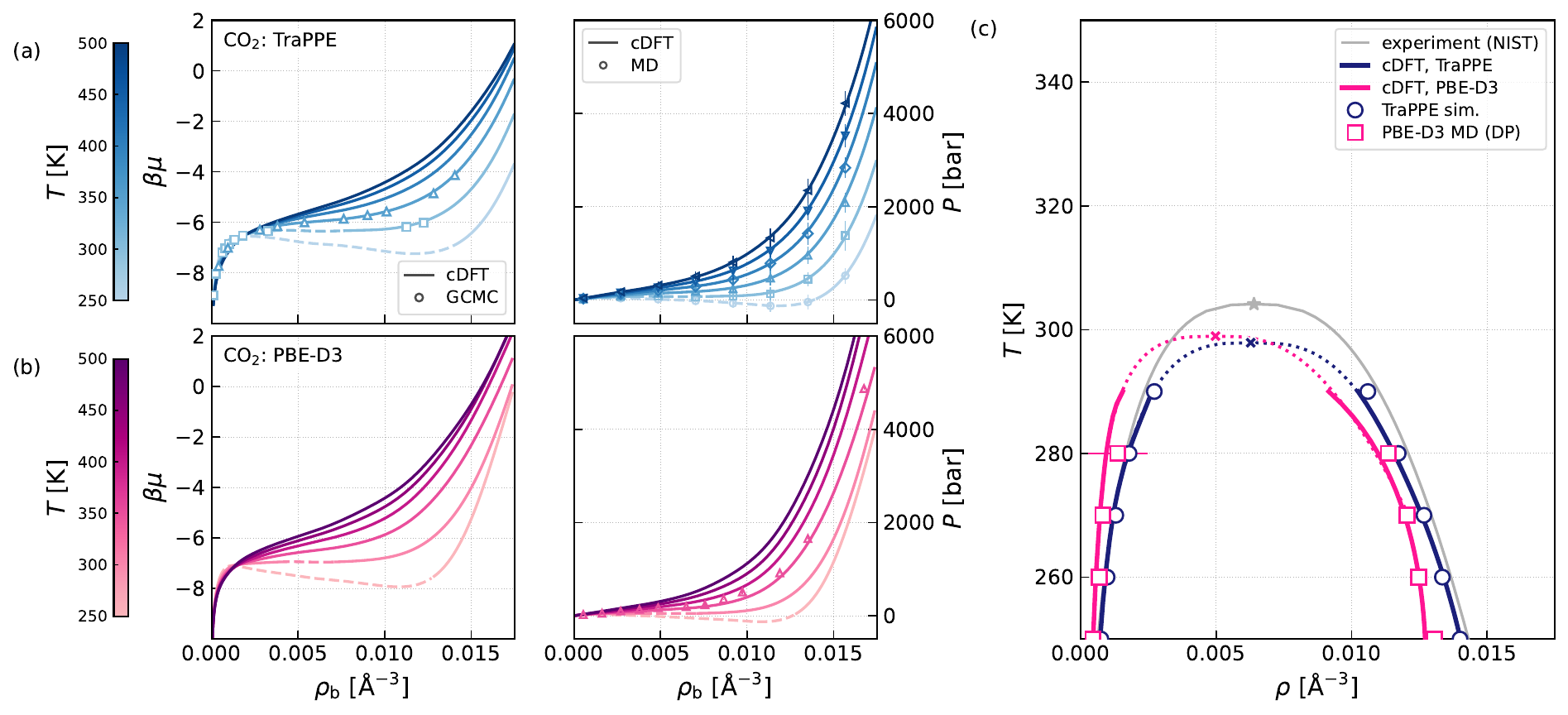}
  \caption{\tbf{Bulk thermodynamics of the CO$_2$ functionals.} (a) Chemical potential and (b) pressure against bulk density for the TraPPE functional from 250 to 500\,K, from its bulk equation of state (lines), with (b) the bulk MD pressures it was trained on (symbols). (c) Vapor--liquid binodal of the TraPPE and PBE-D3 functionals against experiment and direct-coexistence simulation of the corresponding CO$_2$ models (Sec.~\ref{sec:validation}).}
\label{figS2}
\end{figure*}

\textit{Held-out test set.} Fig.~\ref{figS1} compares cDFT with MD for test simulations, none of which was used in training or validation. Over these, $c^{(1)}$ is predicted with an RMSE of 0.25 for TraPPE and 0.35 for PBE-D3, and a single cDFT solve at the simulation's $\beta\mu$ reproduces the number of molecules to 0.9 and 2.1\,\% on average, respectively.

\textit{Bulk thermodynamics.} Fig.~\ref{figS2} shows the bulk equation of state and the vapor--liquid binodal. At each temperature, the coexisting vapor and liquid densities, $\rho_\mrm{v}$ and $\rho_\mrm{l}$, follow from equal chemical potential and pressure in the two phases,
\begin{equation}
  \mu_\mrm{b}(\rho_\mrm{v}) = \mu_\mrm{b}(\rho_\mrm{l}), \qquad P(\rho_\mrm{v}) = P(\rho_\mrm{l}),
  \label{eq:coex}
\end{equation}
with $\mu_\mrm{b}$ and $P$ of the functional as above. Below $T_\mrm{c}$, the equation of state of the uniform fluid has a van der Waals loop. A Maxwell equal-area construction gives $\rho_\mrm{v}$ and $\rho_\mrm{l}$. The TraPPE functional gives $T_\mrm{c} = 298$\,K against 304\,K for experiment \cite{Span1996CO2, NISTWebBook}. For both functionals, the coexisting densities follow those of direct coexistence MD simulations directly with the TraPPE force field and the PBE-D3 MLIP \cite{Bui2026MultiscaleLiquids}.

\section{Interaction energy calculation}
\label{sec:eint}

We compute the interaction energy
\begin{equation}
  E_\mrm{int}(\mbf{r},\bm{\omega}) = E_{\mrm{MOF+CO}_2}(\mbf{r},\bm{\omega}) - E_\mrm{MOF} - E_{\mrm{CO}_2}
  \label{eq:eint}
\end{equation}
for a rigid CO$_2$ molecule in a rigid framework, with the carbon atom on each point $\mbf{r}$ of a grid of spacing 0.5\,\AA{} and the molecular axis along each direction $\bm{\omega}$ of an orientational quadrature. The CO$_2$ geometry is relaxed in the gas phase with each potential. To avoid interactions with periodic images, the framework is replicated until every interplanar distance exceeds 12\,\AA. The external potential applied in cDFT is the Boltzmann average of $E_\mrm{int}(\mbf{r},\bm{\omega})$ over orientations,
\begin{equation}
  U_\mrm{ext}(\mbf{r}) = -k_\mrm{B}T\ln\left\langle \exp\left(-\beta E_\mrm{int}(\mbf{r},\bm{\omega})\right)\right\rangle_{\bm{\omega}},
  \label{eq:uext}
\end{equation}
where the orientational average over the directions $\bm{\omega}$ of the molecular axis is evaluated with a quadrature of $n_\omega$ directions $\bm{\omega}_i$ and weights $w_i$,
\begin{equation}
  \left\langle f \right\rangle_{\bm{\omega}} = \frac{1}{4\pi}\int\!\mrm{d}\bm{\omega}\, f(\bm{\omega}) \approx \sum_{i=1}^{n_\omega} w_i\, f(\bm{\omega}_i), \qquad \sum_{i=1}^{n_\omega} w_i = 1.
  \label{eq:quad}
\end{equation}
$U_\mrm{ext}$ depends on temperature so is therefore recomputed at each $T$ from the same $E_\mrm{int}(\mbf{r},\bm{\omega})$ grid.

For the MLIPs, UMA-s and UMA-m, both v1.1 \cite{Wood2025UMA}, and the eSEN-sm models of Sriram \gammal{} \cite{Fu2025eSEN, sriram2025odac25} are used with their ODAC task head, trained on PBE-D3 calculations, and MACE-DAC \cite{Lim2025DirectAirCapture} with D3 dispersion and Becke--Johnson damping. uMOF-MH and uMOF-POLAR \cite{Inizan2026uMOF} are used with their r$^2$SCAN-D4 head. For UFF \cite{Rappe1992UFF}, the framework atoms interact with TraPPE CO$_2$ through Lennard-Jones potentials only, with Lorentz--Berthelot mixing and cut and shifted at 12\,\AA.

The orientational quadrature is built from orbits of a few seed directions under the point group of the framework, with seeds and weights optimized for polynomial exactness, as for Lebedev grids. The set of orientations is therefore closed under the symmetry of the framework, so that $E_\mrm{int}$ need only be computed at the symmetry-unique grid points. Since CO$_2$ is centrosymmetric, $E_\mrm{int}(\mbf{r},\bm{\omega}) = E_\mrm{int}(\mbf{r},-\bm{\omega})$: $\bm{\omega}$ and $-\bm{\omega}$ are the same orientation, so the directions need only cover a hemisphere, and only spherical harmonics of even degree need be integrated exactly. We use at least 12 directions, and a Gauss--Legendre product rule for frameworks without rotational symmetry. Points within 1.5\,\AA{} of a framework atom are not computed and are excluded from $U_\mrm{ext}$. To reduce the cost further, $E_\mrm{int}$ is first computed on every second grid point along each axis, and then at every grid point near a coarse point whose $U_\mrm{ext}$ at 298\,K lies within 12\,$k_\mrm{B}T$ of the lowest, i.e., in every region that contributes to adsorption. Elsewhere, $E_\mrm{int}$ is interpolated trilinearly from the coarse points for each orientation. For full scans of 10 MOFs, this changes $K_\mrm{H}$ by less than 0.01\,\%. The cost of the scans is compared in Sec.~\ref{sec:cost}.

$U_\mrm{ext}$ is then projected onto the 0.5\,\AA{} grid of the cDFT by trilinear interpolation of $\exp(-\beta U_\mrm{ext})$, in the smallest commensurate supercell with a strain below 3\,\%. Voxels where every orientation overlaps the framework, or where $\beta U_\mrm{ext} > 60$, are excluded and define the accessible volume $V_\mrm{acc}$.

\begin{figure*}[h]
  \includegraphics[width=\linewidth]{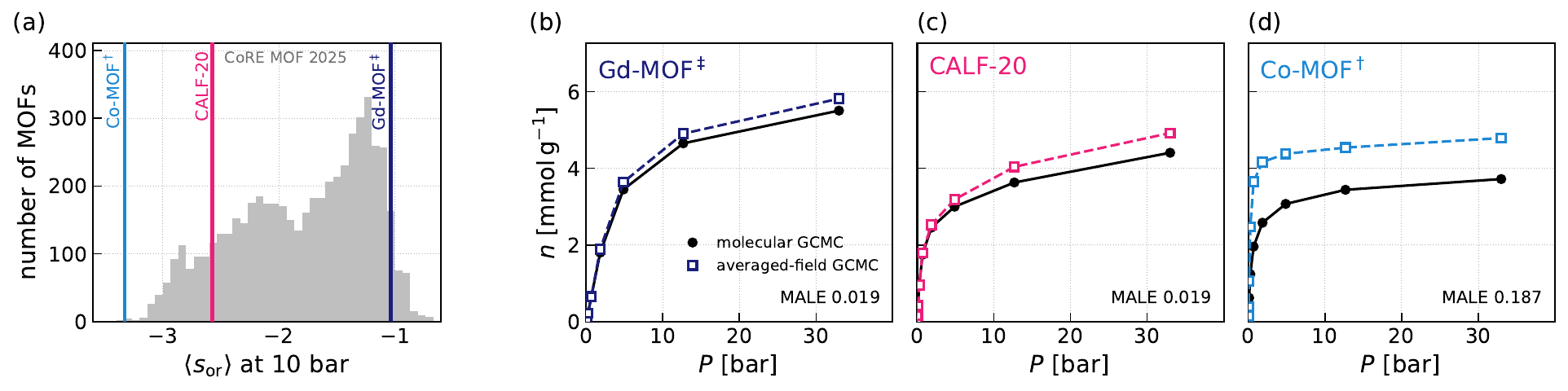}
  \caption{\tbf{Validity of the orientational average of the external potential.} UFF framework and TraPPE CO$_2$ at 298\,K. (a) $\langle s_\mrm{or}\rangle$ at 10\,bar (Eq.~\eqref{eq:sor}) across the CoRE MOF 2025 database (gray), and of the three MOFs of (b)--(d) (lines). (b)--(d) Uptake from molecular (circles) and averaged-field GCMC (squares) at the same chemical potential, with the MALE (Eq.~\eqref{eq:male}) of the latter. CoRE MOF 2025 structures $^\ddagger$2018[Gd][fcu]3[ASR]1 and $^\dagger$2014[Co][nan]3[ASR]37.}
\label{figSorient}
\end{figure*}

\textit{Validity of the orientational average.} Eq.~\eqref{eq:uext} is exact in the dilute limit. At finite loading, the cDFT describes CO$_2$ by the density of its carbon in $U_\mrm{ext}$, as in its training, and so neglects how the orientations imposed by the framework change the CO$_2$--CO$_2$ interactions. We measure how strongly the framework locks CO$_2$ by the orientational entropy
\begin{equation}
  s_\mrm{or}(\mbf{r}) = \beta\left[U_\mrm{min}(\mbf{r}) - U_\mrm{ext}(\mbf{r})\right] \le 0,
  \label{eq:sor}
\end{equation}
with $U_\mrm{min}$ the lowest $E_\mrm{int}$ over the quadrature, so that $\exp(s_\mrm{or})$ is the fraction of orientations available. Its average $\langle s_\mrm{or}\rangle$ over where CO$_2$ sits at 10\,bar, from a local Langmuir density in $U_\mrm{ext}$, ranges from $-3.0$ to $-0.9$ across the database with UFF (Fig.~\ref{figSorient}a).

To test the approximation, we compare GCMC simulations of rigid TraPPE CO$_2$ with UFF acting on every site of CO$_2$ (molecular, RASPA3 \cite{Ran2024RASPA3}) and with $U_\mrm{ext}$ acting on the carbon only (averaged field, our own code), with the same CO$_2$--CO$_2$ interactions and at the fugacities of the cDFT chemical potentials, in a weakly locked Gd-MOF, CALF-20 and the most locked porous MOF of the database, up to 40\,bar (Fig.~\ref{figSorient}b--d). Both use Lennard-Jones interactions cut and shifted at 15\,\AA{} and Ewald electrostatics. In the Gd-MOF, which leaves CO$_2$ nearly free to rotate, the two agree to within 6\,\% at every pressure, and the 5\,\% shortfall of the averaged field at low loading comes from the 24-direction quadrature, which underestimates its $K_\mrm{H}$ by as much. In CALF-20, they agree to within 3\,\% up to 2\,bar, and the averaged field overestimates the uptake by up to 12\,\% at 33\,bar, since freely rotating CO$_2$ packs more densely than CO$_2$ aligned by the framework. In the Co-MOF, the averaged field overestimates the uptake by up to a factor of 2 at intermediate loading. About a quarter of this error comes from the 24-direction quadrature, which overestimates $K_\mrm{H}$ by 29\,\% where the allowed cone is narrow, and the rest from the orientational average itself. The approximation therefore holds for MOFs that lock CO$_2$ as strongly as CALF-20, which covers 88\,\% of the database, while the most strongly locked MOFs can be overestimated substantially.

\section{CO$_2$ in MOFs: cDFT against simulation}
\label{sec:mofsim}

\begin{figure*}[h]
  \includegraphics[width=0.7\linewidth]{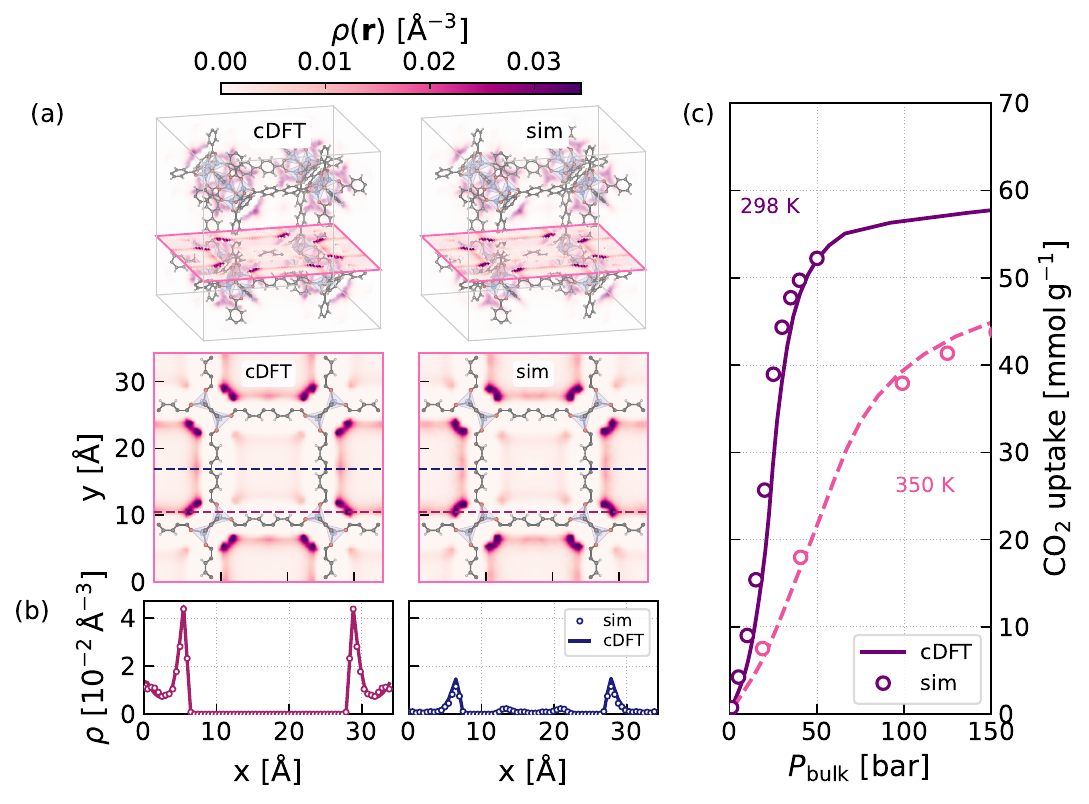}
  \caption{\tbf{CO$_2$ in Zn-IRMOF-10, cDFT against simulation.} As the adsorption properties of Fig.~1 of the main text, for Zn-IRMOF-10 (TraPPE CO$_2$ functional, force-field external potential). (a) CO$_2$ density at 298\,K and 74 molecules per cell, cDFT (left) and NVT MD (sim, right): the 3D density with the plane of the slice highlighted, and the density in the slice through the deepest $U_\mrm{ext}$ site. (b) A 1D cut along $x$ below each slice (dashed on the maps), cDFT against simulation. (c) Absolute uptake against bulk pressure at 298\,K (solid) and 350\,K (dashed), cDFT against simulation (circles), as described in Sec.~\ref{sec:mofsim}.}
\label{figS3}
\end{figure*}

\begin{figure*}[h]
  \includegraphics[width=0.95\linewidth]{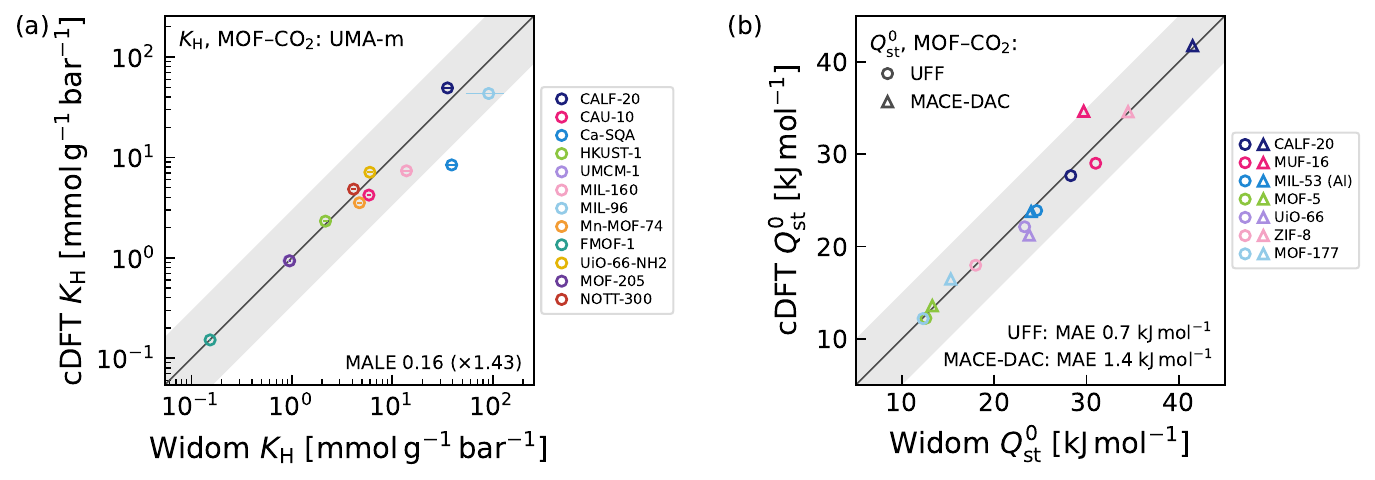}
  \caption{\tbf{The cDFT dilute limit against Widom insertion with the same MOF--CO$_2$ potential.} Left: CO$_2$ Henry constant $K_\mrm{H}$ at 298\,K with UMA-m, against the Widom-insertion values of Sriram \gammal{} \cite{sriram2025odac25}, for the MOFs of their experimental set whose structure matches the one used here. The band marks agreement within a factor of 3. Right: zero-loading heat of adsorption $Q^0_\mrm{st}$ with UFF (circles) and MACE-DAC (triangles), against the Widom-insertion values of Lim \gammal{} \cite{Lim2025DirectAirCapture}. The band marks $\pm5$\,kJ\,mol$^{-1}$.}
\label{figS4}
\end{figure*}

\textit{Dilute limit.} The external potential alone gives the Henry constant,
\begin{equation}
  K_\mrm{H} = \frac{\beta}{V_\mrm{cell}}\int\!\mrm{d}\mbf{r}\,\exp\left(-\beta U_\mrm{ext}(\mbf{r})\right),
  \label{eq:kh}
\end{equation}
and the zero-loading heat of adsorption $Q^0_\mrm{st}$, with no cDFT solve. For the same MOF--CO$_2$ potentials, both agree with the Widom-insertion values of Sriram \gammal{} \cite{sriram2025odac25} and of Lim \gammal{} \cite{Lim2025DirectAirCapture}, as shown in Fig.~\ref{figS4}.

\textit{Adsorption in IRMOF-1 and IRMOF-10.} For the comparisons of Figs.~1 and \ref{figS3}, the framework is described by the force field of Ref.~\onlinecite{Borycz2016CO2Adsorption}, UFF Lennard-Jones parameters with CM5 charges, with TraPPE CO$_2$, and the same interactions are used for $E_\mrm{int}(\mbf{r},\bm{\omega})$ in cDFT. The same reference provides Gibbs-ensemble Monte Carlo (GEMC) isotherms of both frameworks at 298\,K, against which we compare. MD simulations were performed with LAMMPS \cite{Thompson2022LAMMPS} in the NVT ensemble at 298\,K with the settings of the training simulations: for IRMOF-1, 496 molecules in a $2\times2\times2$ supercell for 0.7\,ns, with the density folded onto one unit cell, and for IRMOF-10, 74 molecules per cell for 150\,ns. GCMC simulations were performed with DL\_MONTE \cite{Brukhno2021DLMONTE} in a $2\times2\times2$ supercell of IRMOF-1 at 350 and 400\,K, and of IRMOF-10 at 350\,K, with Ewald electrostatics, 20\,\% translation, 20\,\% rotation and 60\,\% insertion/deletion moves, and $5\times10^6$ moves per state point, of which the first $2\times10^6$ were discarded. The pressure of each state point follows from its bulk density through the bulk equation of state of TraPPE CO$_2$ from MD. Uptakes in these comparisons are absolute.

\section{Compute cost comparison}
\label{sec:cost}

\begin{figure*}[h]
  \includegraphics[width=\linewidth]{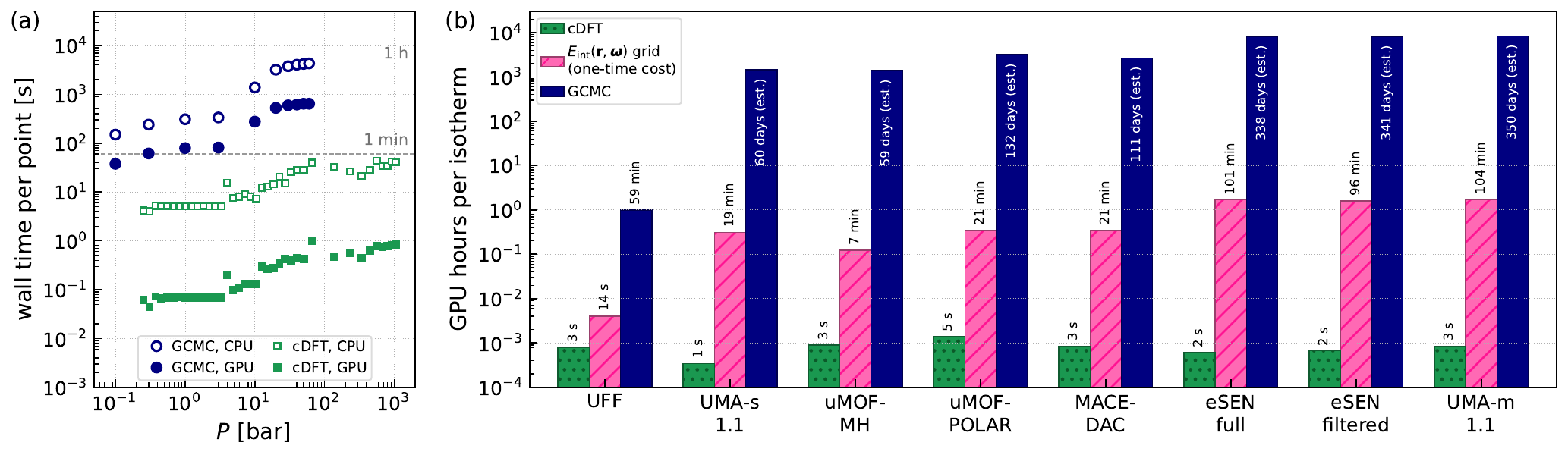}
  \caption{\tbf{Compute cost of cDFT against GCMC.} CO$_2$ in IRMOF-1 at 298\,K, as described in Sec.~\ref{sec:cost}. (a) Wall time per isotherm point. (b) Wall time of an isotherm on one GPU for each MOF--CO$_2$ potential of Fig.~2 of the main text: the GCMC pressures of (a), and for cDFT the nearest points of (a), with the $E_\mrm{int}(\mbf{r},\bm{\omega})$ grid (hatched) needed once per MOF. With UFF, its time is estimated from its energy evaluations run on the GPU. With an MLIP, the grid of $1.7\times10^4$ single points dominates. GCMC with an MLIP: a rough estimate (see text).}
\label{figS5}
\end{figure*}

We compare the wall time of an isotherm of CO$_2$ in IRMOF-1 at 298\,K, with UFF for the MOF--CO$_2$ interaction and TraPPE CO$_2$, on one NVIDIA H100 GPU or one AMD EPYC 9554 CPU core in Fig.~\ref{figS5}. GCMC simulations use gRASPA \cite{Li2024gRASPA} on the GPU and RASPA3 \cite{Ran2024RASPA3} on the CPU, with $1$--$8\times10^6$ production moves per pressure. Each cDFT solve takes seconds, so with an MLIP the cost of an isotherm is set by the $E_\mrm{int}(\mbf{r},\bm{\omega})$ grid, computed once per MOF: from 7\,min with uMOF-MH to 1.7\,h with UMA-m. For reference, we also give a rough estimate for GCMC with an MLIP, from the wall time per step of MLIP-MC \cite{Edwards2025MLIPMC}, which evaluates the full-system MLIP energy for each trial move, times the number of production moves used with UFF.

\section{Experimental comparison}
\label{sec:expt}

\begin{figure*}[h]
  \includegraphics[width=0.85\linewidth]{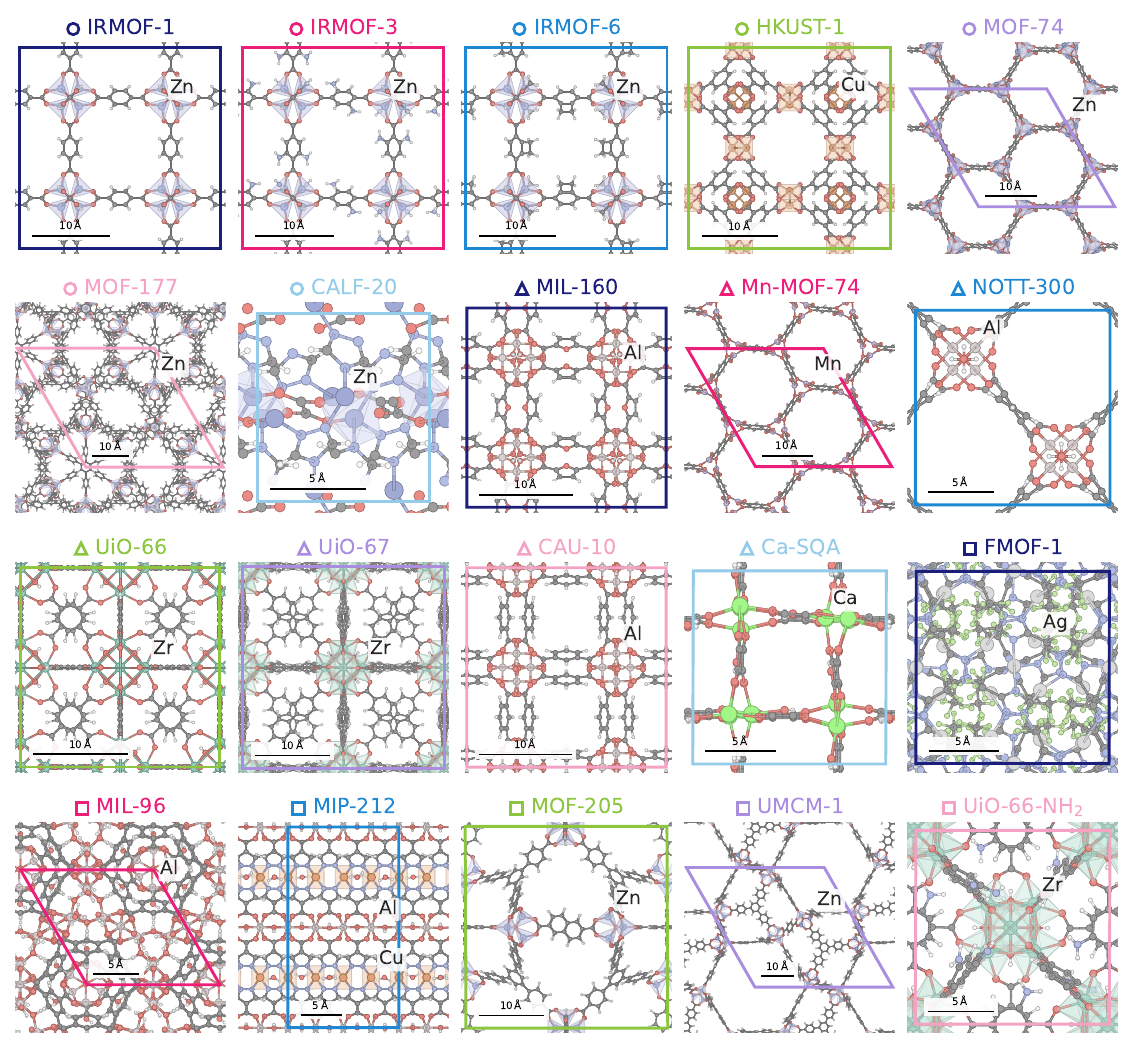}
  \caption{\tbf{The frameworks of the experimental comparison.} Unit cells of the 20 MOFs compared with experiment, each viewed down its $c$ axis with one cell outlined and its own 10\,\AA\ scale bar (5\,\AA\ for cells narrower than 20\,\AA). Each name is in the color and symbol used for that MOF throughout. Centered lattices are shown in their conventional cells. CALF-20, whose monoclinic cell stacks offset layers in projection, is shown as the 7\,\AA\ slab around the plane through its deepest binding site, as in Fig.~3(b) of the main text.}
\label{figS6}
\end{figure*}

\begin{figure*}[h]
  \includegraphics[width=\linewidth]{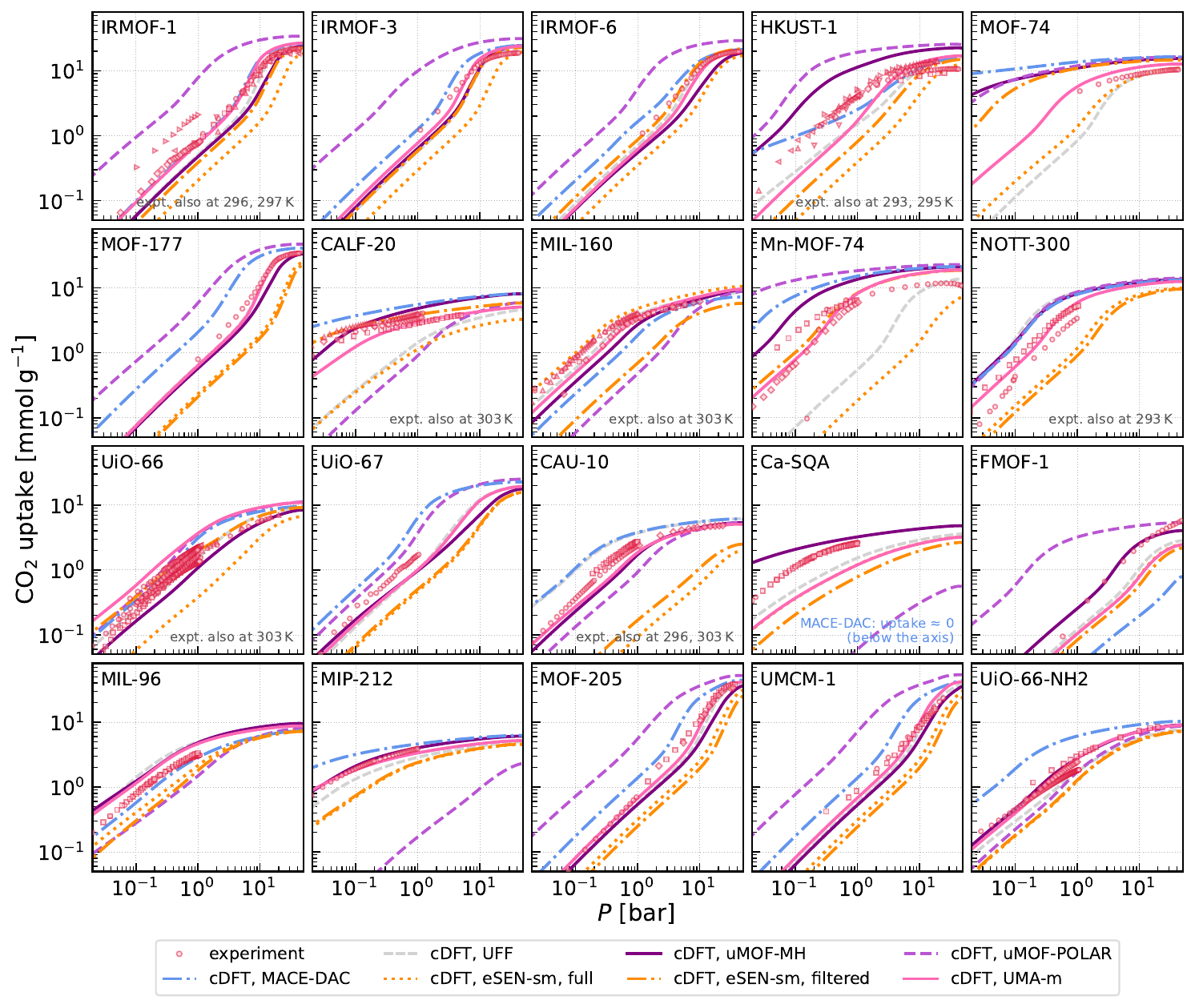}
  \caption{\tbf{Experimental and cDFT isotherms of the 20 MOFs.} CO$_2$ uptake at 298\,K on a log pressure axis: experiment (circles) and cDFT with the TraPPE CO$_2$ functional and the external potential from UFF, uMOF-MH, uMOF-POLAR, MACE-DAC, eSEN-sm (full and filtered) and UMA-m (lines). Where a MOF was also measured at other temperatures near 298\,K, these are noted in its panel.}
\label{figS7}
\end{figure*}

\begin{figure*}[h]
  \includegraphics[width=\linewidth]{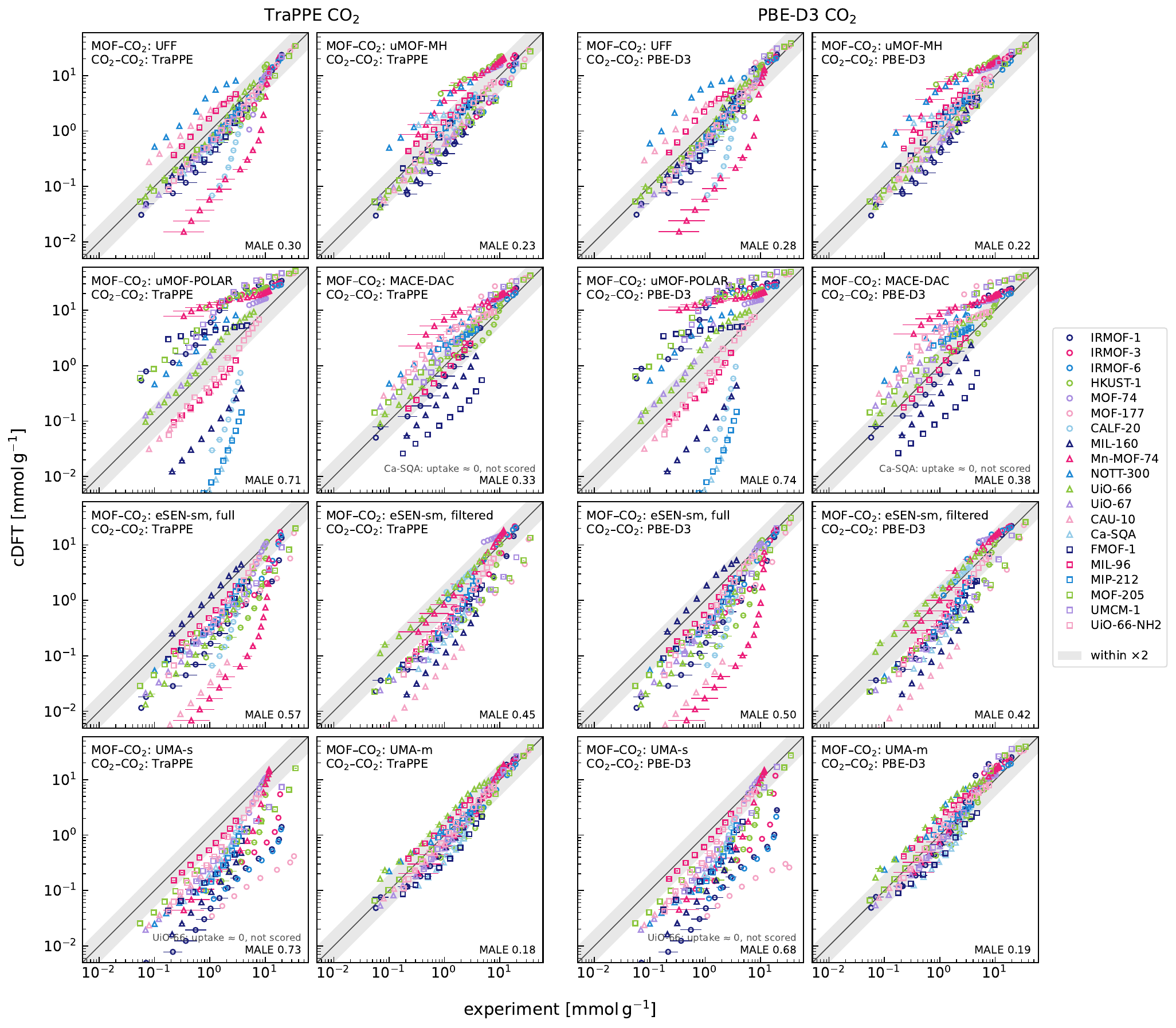}
  \caption{\tbf{Parity of cDFT and experimental CO$_2$ uptake for each MOF--CO$_2$ external potential and CO$_2$--CO$_2$ functional.} The left block uses the TraPPE CO$_2$ functional and the right block the PBE-D3 functional. Within each, the panels give the potential used for the $E_\mrm{int}(\mbf{r},\bm{\omega})$ grid (UFF, uMOF-MH, uMOF-POLAR, MACE-DAC, eSEN-sm full and filtered, UMA-s and UMA-m), in the same place in both blocks. Every point is one experimental isotherm point of one MOF of Fig.~\ref{figS6}, and the shaded band marks agreement within a factor of 2. Each panel gives the mean absolute log error (MALE), the corresponding factor and the number of MOFs scored.}
\label{figS8}
\end{figure*}

For each of the 20 MOFs of Fig.~\ref{figS6}, the experimental isotherms within 1\,K of 298\,K \cite{Millward2005CarbonDioxideStorage, Charalambous2024CarbonCapture, cavka2014high} are interpolated onto a common grid of pressures $P_i$, five per decade from 0.01\,bar, and averaged in $\log_{10}$ into one consensus uptake $N^\mrm{expt}_m(P_i)$ of MOF $m$. The cDFT excess uptake $N^\mrm{cDFT}_m(P_i)$ is interpolated onto the same grid, and Fig.~\ref{figS7} compares the two. We measure their agreement by the mean absolute log error,
\begin{equation}
  \mrm{MALE}_m = \frac{1}{n_m}\sum_{i=1}^{n_m}\left|\log_{10}\frac{N^\mrm{cDFT}_m(P_i)}{N^\mrm{expt}_m(P_i)}\right|, \qquad \mrm{MALE} = \frac{1}{M}\sum_{m=1}^{M}\mrm{MALE}_m,
  \label{eq:male}
\end{equation}
where $n_m$ is the number of grid points with both uptakes and $N^\mrm{expt}_m \geq 0.01$\,mmol\,g$^{-1}$, and $M$ the number of MOFs. Fig.~\ref{figS8} gives it for each MOF--CO$_2$ potential and CO$_2$ functional. A log error weighs a miss by a given factor equally at low and high uptake, and averaging per MOF first lets every MOF count once, however many isotherms it has. $10^\mrm{MALE}$ is the typical factor by which the cDFT and experimental uptakes differ.
Table~\ref{tab:male_mof} breaks the MALE down by MOF.

\begin{table*}[h]
\small
\centering
\setlength{\tabcolsep}{4pt}
\renewcommand{\arraystretch}{1.1}
\caption{\tbf{Mean absolute log error by MOF.} $\mrm{MALE}_m$ of Eq.~\ref{eq:male} for each MOF of Fig.~\ref{figS6} and each MOF--CO$_2$ potential, with the TraPPE (top) and PBE-D3 (bottom) CO$_2$ functional. The rows marked ``all'' give the MALE of Fig.~2 of the main text. A dash marks a MOF not scored with that potential.}
\label{tab:male_mof}
\begin{tabular}{lcccccccc}
\hline
\hline
{\bfseries\boldmath MOF} & {\bfseries\boldmath UFF} & {\bfseries\boldmath \begin{tabular}[c]{@{}c@{}}uMOF-\\MH\end{tabular}} & {\bfseries\boldmath \begin{tabular}[c]{@{}c@{}}uMOF-\\POLAR\end{tabular}} & {\bfseries\boldmath \begin{tabular}[c]{@{}c@{}}MACE-\\DAC\end{tabular}} & {\bfseries\boldmath \begin{tabular}[c]{@{}c@{}}eSEN-sm\\full\end{tabular}} & {\bfseries\boldmath \begin{tabular}[c]{@{}c@{}}eSEN-sm\\filtered\end{tabular}} & {\bfseries\boldmath UMA-s} & {\bfseries\boldmath UMA-m} \\
\hline
\multicolumn{9}{l}{\textit{TraPPE CO$_2$ functional}} \\
IRMOF-1 & 0.29 & 0.34 & 0.68 & 0.16 & 0.72 & 0.42 & 1.32 & 0.17 \\
IRMOF-3 & 0.21 & 0.19 & 0.52 & 0.13 & 0.50 & 0.20 & 0.99 & 0.12 \\
IRMOF-6 & 0.08 & 0.26 & 0.51 & 0.13 & 0.41 & 0.05 & 1.30 & 0.13 \\
HKUST-1 & 0.27 & 0.42 & 0.60 & 0.13 & 0.70 & 0.44 & 0.69 & 0.25 \\
Zn-MOF-74 & 0.20 & 0.25 & 0.28 & 0.30 & 0.14 & 0.23 & 0.09 & 0.10 \\
MOF-177 & 0.10 & 0.21 & 0.60 & 0.37 & 0.62 & 0.69 & 1.69 & 0.11 \\
CALF-20 & 0.87 & 0.08 & 1.16 & 0.23 & 0.91 & 0.05 & 0.58 & 0.25 \\
MIL-160 & 0.13 & 0.35 & 1.14 & 0.53 & 0.08 & 0.94 & 1.03 & 0.21 \\
Mn-MOF-74 & 0.84 & 0.37 & 0.66 & 0.50 & 1.39 & 0.10 & 0.50 & 0.15 \\
NOTT-300 & 0.62 & 0.57 & 0.57 & 0.55 & 0.42 & 0.52 & 0.73 & 0.23 \\
UiO-66 & 0.08 & 0.19 & 0.20 & 0.14 & 0.60 & 0.10 & -- & 0.30 \\
UiO-67 & 0.23 & 0.15 & 0.21 & 0.38 & 0.55 & 0.51 & 0.59 & 0.19 \\
CAU-10 & 0.41 & 0.19 & 0.44 & 0.43 & 1.50 & 1.21 & 0.56 & 0.12 \\
Ca-SQA & 0.32 & 0.28 & 2.11 & -- & 0.45 & 0.70 & 0.66 & 0.45 \\
FMOF-1 & 0.33 & 0.04 & 0.58 & 0.98 & 0.41 & 0.52 & 0.71 & 0.42 \\
MIL-96 & 0.23 & 0.21 & 0.42 & 0.11 & 0.26 & 0.39 & 0.22 & 0.18 \\
MIP-212 & 0.16 & 0.03 & 1.83 & 0.22 & 0.39 & 0.37 & 0.71 & 0.04 \\
MOF-205 & 0.14 & 0.19 & 0.71 & 0.27 & 0.43 & 0.55 & 0.49 & 0.08 \\
UMCM-1 & 0.18 & 0.24 & 0.75 & 0.24 & 0.49 & 0.61 & 0.55 & 0.13 \\
UiO-66-NH$_2$ & 0.19 & 0.03 & 0.25 & 0.39 & 0.35 & 0.35 & 0.38 & 0.05 \\
all & 0.30 & 0.23 & 0.71 & 0.33 & 0.57 & 0.45 & 0.73 & 0.18 \\
\hline
\multicolumn{9}{l}{\textit{PBE-D3 CO$_2$ functional}} \\
IRMOF-1 & 0.28 & 0.29 & 0.77 & 0.20 & 0.64 & 0.36 & 1.28 & 0.20 \\
IRMOF-3 & 0.15 & 0.15 & 0.56 & 0.27 & 0.35 & 0.15 & 0.86 & 0.16 \\
IRMOF-6 & 0.11 & 0.12 & 0.57 & 0.24 & 0.27 & 0.14 & 1.25 & 0.10 \\
HKUST-1 & 0.20 & 0.47 & 0.61 & 0.10 & 0.62 & 0.36 & 0.61 & 0.19 \\
Zn-MOF-74 & 0.09 & 0.24 & 0.27 & 0.28 & 0.06 & 0.22 & 0.04 & 0.13 \\
MOF-177 & 0.12 & 0.09 & 0.69 & 0.51 & 0.48 & 0.54 & 1.72 & 0.12 \\
CALF-20 & 0.82 & 0.10 & 1.13 & 0.27 & 0.87 & 0.10 & 0.53 & 0.19 \\
MIL-160 & 0.08 & 0.26 & 1.12 & 0.46 & 0.19 & 0.92 & 1.01 & 0.12 \\
Mn-MOF-74 & 0.77 & 0.43 & 0.67 & 0.54 & 1.33 & 0.14 & 0.44 & 0.17 \\
NOTT-300 & 0.74 & 0.70 & 0.71 & 0.67 & 0.37 & 0.49 & 0.71 & 0.35 \\
UiO-66 & 0.09 & 0.14 & 0.29 & 0.21 & 0.54 & 0.18 & -- & 0.37 \\
UiO-67 & 0.20 & 0.13 & 0.30 & 0.51 & 0.54 & 0.50 & 0.58 & 0.17 \\
CAU-10 & 0.50 & 0.12 & 0.40 & 0.52 & 1.50 & 1.20 & 0.53 & 0.05 \\
Ca-SQA & 0.30 & 0.31 & 2.11 & -- & 0.44 & 0.69 & 0.65 & 0.43 \\
FMOF-1 & 0.24 & 0.12 & 0.60 & 0.94 & 0.33 & 0.44 & 0.64 & 0.33 \\
MIL-96 & 0.32 & 0.30 & 0.36 & 0.05 & 0.20 & 0.34 & 0.15 & 0.27 \\
MIP-212 & 0.10 & 0.10 & 1.83 & 0.27 & 0.34 & 0.32 & 0.68 & 0.04 \\
MOF-205 & 0.11 & 0.10 & 0.81 & 0.37 & 0.34 & 0.46 & 0.40 & 0.09 \\
UMCM-1 & 0.19 & 0.19 & 0.89 & 0.38 & 0.38 & 0.51 & 0.44 & 0.18 \\
UiO-66-NH$_2$ & 0.13 & 0.09 & 0.17 & 0.45 & 0.28 & 0.28 & 0.32 & 0.08 \\
all & 0.28 & 0.22 & 0.74 & 0.38 & 0.50 & 0.42 & 0.68 & 0.19 \\
\hline
\hline
\end{tabular}
\end{table*}

With an MLIP, the cost of a cDFT isotherm is that of the $E_\mrm{int}(\mbf{r},\bm{\omega})$ grid (Sec.~\ref{sec:cost}). Among the MLIPs, only UMA-m and uMOF-MH reproduce the experiments to within a factor of 2 on average. UMA-m is the most accurate, but uMOF-MH comes within 0.03 of its MALE at an order of magnitude less cost, which makes the two the potentials of choice for the database, where the grid of every MOF has to be computed. UFF has a similar average error, but most of these MOFs bind CO$_2$ by physisorption alone, and UFF misses the uptake of Mn-MOF-74, CALF-20 and NOTT-300, which bind CO$_2$ strongly, by factors of 5 to 7 (Table~\ref{tab:male_mof}).

\begin{figure*}[h]
  \centering
  \includegraphics[width=0.5\linewidth]{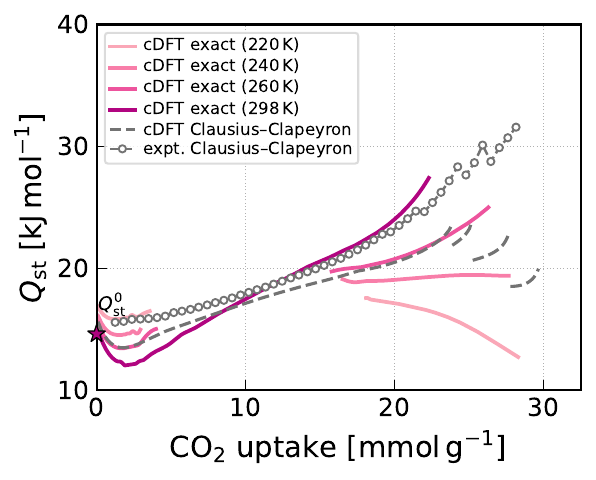}
  \caption{\tbf{Isosteric heat of IRMOF-1 at several temperatures.} $Q_\mrm{st}(n)$ from the cDFT isotherms (UMA-m external potential, PBE-D3 CO$_2$ functional) from the fugacity, without the ideal-gas assumption (solid) at 220, 240, 260 and 298\,K, and from the integrated Clausius--Clapeyron form over 220--310\,K (dashed), against the experimental values of Simmons et al.\ \cite{simmons2011carbon} from the same procedure (circles). The star marks the zero-loading value $Q^0_\mrm{st}$ from the potential energy scan. Each curve is broken across the pore-filling step of its isotherms, where no loading is computed.}
\label{figSqst}
\end{figure*}

Fig.~\ref{figSqst} extends Fig.~3(d) of the main text, which shows $Q_\mrm{st}(n)$ at 298\,K only, to lower temperatures. At each temperature $T_0$, the derivative of $\ln f$ in $1/T$ at fixed $n$ is taken from the isotherms at $T_0$ and $T_0 \pm 5$\,K, or from the neighboring temperatures where one of these is missing.

\clearpage

\section{Screened structures}
\label{sec:db}

We screen the 5,591 computation-ready, all-solvent-removed (CR-ASR) structures of the CoRE MOF 2025 database \cite{Zhao2025CoREMOFDB}. With UFF, MACE-DAC and uMOF-MH, we attempted every structure and obtained the $E_\mrm{int}$ grid of 5,451, 5,486 and 5,207 of them, respectively. The scans of the rest failed, because of overlapping atoms or other artifacts in the structure. For UMA-m, the most expensive potential, we computed 4,004 MOFs, 72\,\% of the database, mostly those with fewer atoms per unit cell, which are the least costly to scan: 87\,\% of the MOFs with fewer than 200 atoms per unit cell, against 17\,\% of those with 600 to 1,000. 

\section{Isotherm classification}
\label{sec:class}

\begin{figure*}[h]
  \includegraphics[width=\linewidth]{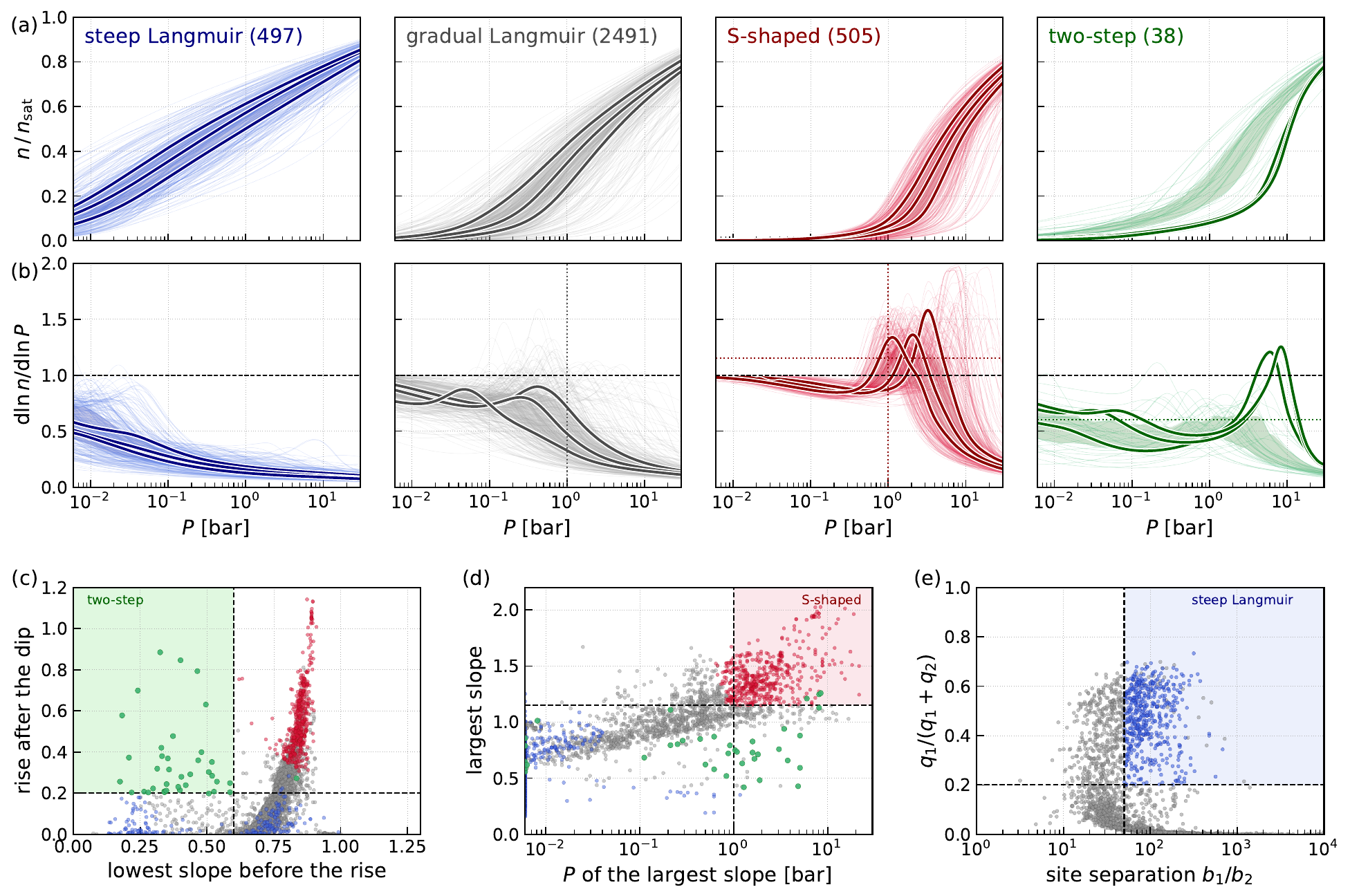}
  \caption{\tbf{How the isotherm classes of Fig.~4 of the main text are assigned.} cDFT isotherms at 298\,K across the CoRE MOF 2025 database, classified as described in Sec.~\ref{sec:class}. (a) The isotherms, each normalized to its saturation uptake $n_\mathrm{sat}$, for up to 300 MOFs per class, with their interquartile range (shaded) and three examples (bold). (b) Their slope, with Henry's law at 1 (dashed). (c)--(e) The statistic of each test, in the order applied: (c) two-step, the lowest slope before the later peak against the rise from the dip to the peak. (d) S-shaped, the largest slope against the pressure at which it occurs, where an upturn below 1\,bar, as in the gray MOFs above the dashed line on its left, falls outside the test. (e) Steep and gradual Langmuir, from the dual-site fit. Dashed lines mark the thresholds.}
\label{figS10}
\end{figure*}

We classify each cDFT isotherm at 298\,K by its slope $\mathrm{d}\ln n/\mathrm{d}\ln P$ from $6\times10^{-3}$ to 30\,bar, smoothed with a Savitzky--Golay filter over approximately 0.3 decades in $P$. MOFs with no uptake, below 0.1\,mmol\,g$^{-1}$ at 30\,bar, and those with $K_\mrm{H} > 10^4$\,mmol\,g$^{-1}$\,bar$^{-1}$, beyond physisorption, are not classified. The rest are assigned by three tests in turn. A MOF is two-step where the slope dips below 0.6 and then rises again by at least 0.2 to a later peak, and S-shaped where it exceeds 1.15 at some $P\geq1$\,bar. The remaining, Langmuir-like MOFs are fitted with the dual-site Langmuir model \cite{Langmuir1918Adsorption, Mathias1996DualSite},
\begin{equation}
  n(P) = \frac{q_1 b_1 P}{1 + b_1 P} + \frac{q_2 b_2 P}{1 + b_2 P},
  \label{eq:dsl}
\end{equation}
with capacities $q_1, q_2$ and affinities $b_1 > b_2$, by least squares in $\ln n$ over the same pressure window, and with its single-site form, $q_2 = 0$. A MOF is steep Langmuir where the strong site binds more than 50 times as strongly as the other, $b_1/b_2 > 50$, i.e., by approximately 10\,kJ\,mol$^{-1}$, holds more than 20\,\% of the capacity, $q_1/(q_1+q_2) > 0.2$, and the dual-site fit halves the root-mean-square residual in $\ln n$ of the single-site one. Otherwise, it is gradual Langmuir. The two-step test comes first, since a second rise can also exceed the S-shaped threshold. Two-step and S-shaped MOFs stand apart from the rest, whereas steep and gradual Langmuir MOFs form a continuum, such that their shares depend on the cut. Fig.~\ref{figS10} shows the isotherms and slopes of each class, and the statistic of each test.

\begin{figure}[h]
  \includegraphics[width=0.6\linewidth]{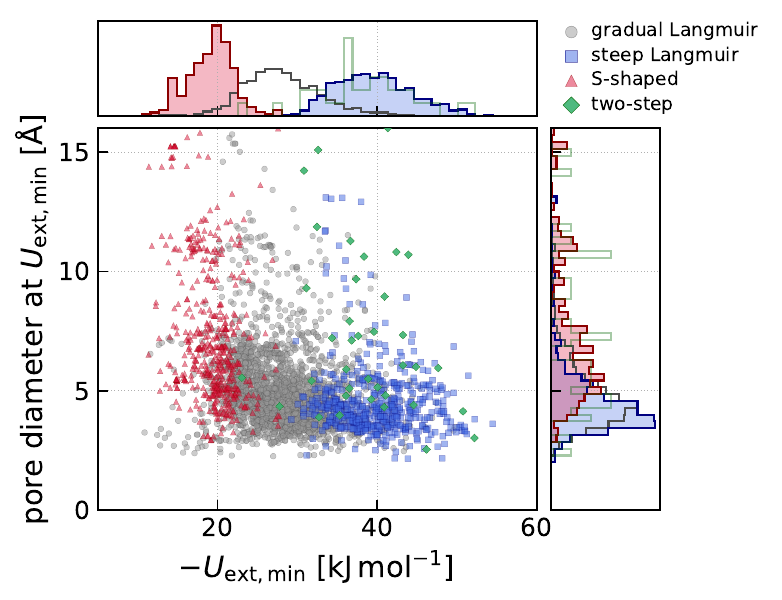}
  \caption{\tbf{Pore size and binding site strength of the four isotherm classes.} The diameter of the pore that holds the deepest $U_\mrm{ext}$ of each MOF, the largest empty sphere that contains that point, against its depth, at 298\,K with UMA-m, colored by the class of the cDFT isotherm (TraPPE CO$_2$ functional, Sec.~\ref{sec:class}). The margins show the distribution of each class along each axis, normalized per class.}
\label{figSporeclass}
\end{figure}

The four classes follow from two properties of the framework, the strength of its strongest binding site and the size of the pore that holds it, as shown in Fig.~\ref{figSporeclass}. We measure that size as the diameter of the largest empty sphere, with van der Waals radii, that contains the deepest point of $U_\mrm{ext}$. Its maximum over the grid reproduces the largest cavity diameter of the CoRE MOF 2025 database \cite{Zhao2025CoREMOFDB} to within 0.2\,\AA. Steep Langmuir MOFs hold strong sites, with a median depth of 40\,kJ\,mol$^{-1}$, in narrow pores, with a median diameter of 4.2\,\AA. Their strong sites fill first and hold a large share of the capacity. Gradual Langmuir MOFs have similarly narrow pores, 4.6\,\AA, but weaker sites, 28\,kJ\,mol$^{-1}$, so that their sites fill over a wider range of pressure. S-shaped MOFs have the weakest sites, 20\,kJ\,mol$^{-1}$, in wider pores, 6.6\,\AA. CO$_2$ is then bound by the framework too weakly to fill the pores until CO$_2$--CO$_2$ attraction takes over, which steepens the isotherm. Two-step MOFs hold sites as strong as those of steep Langmuir MOFs, 38\,kJ\,mol$^{-1}$, in pores of 6.0\,\AA, but their largest cavities, 9.6\,\AA, are as wide as those of S-shaped MOFs, and in 38\,\% of them the strongest site lies in a pocket narrower than 70\,\% of the largest cavity. The strong sites fill first, and the large pores later, cooperatively, as in S-shaped MOFs.

\begin{figure*}[h]
  \includegraphics[width=\linewidth]{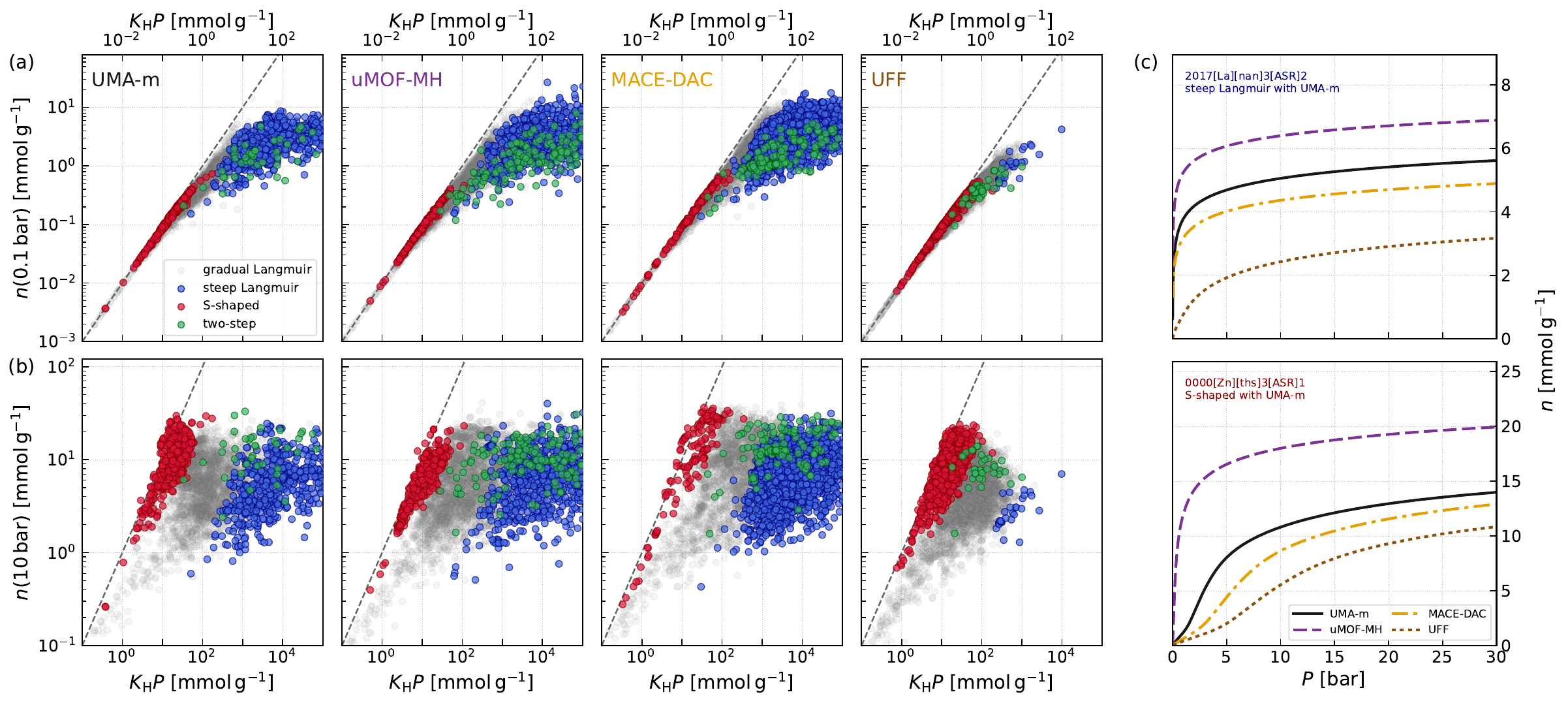}
  \caption{\tbf{The isotherm classes with the UMA-m, uMOF-MH, MACE-DAC and UFF external potentials.} cDFT isotherms at 298\,K (TraPPE CO$_2$ functional) of the MOFs of the CoRE MOF 2025 database with an isotherm under each potential, each MOF classified as in Fig.~\ref{figS10} with that potential. (a) Uptake at 0.1\,bar and (b) at 10\,bar against $K_\mrm{H}P$, with $K_\mrm{H}$ the Henry constant, as Fig.~4(a) of the main text, with Henry's law $n = K_\mrm{H}P$ (dashed). MOFs that are not classified (Sec.~\ref{sec:class}) are not drawn. (c) The isotherms with each potential of two representative MOFs, steep Langmuir and S-shaped with UMA-m.}
\label{figS11}
\end{figure*}

Fig.~\ref{figS11} applies the same classification with the uMOF-MH, MACE-DAC and UFF potentials, each to every MOF it was run for, and the shares of the classes depend strongly on the potential. Without strong binding sites, UFF finds almost no steep Langmuir MOFs, against 13\,\% of the database with UMA-m, and CO$_2$--CO$_2$ attraction instead shapes nearly twice as many isotherms into S-shaped ones. MACE-DAC overbinds instead: it makes nearly three times as many MOFs steep Langmuir, places nearly twice as many beyond physisorption and raises the uptake at every pressure. uMOF-MH finds steep Langmuir MOFs about as often as UMA-m, but binds more strongly at the strongest sites, placing nearly three times as many MOFs beyond physisorption, and finds fewer than half as many S-shaped ones. The representative MOFs of Fig.~\ref{figS11}(c) show the same trends: for the steep Langmuir MOF, UFF falls well below UMA-m, and for the S-shaped one, MACE-DAC lies well above it, whereas UFF follows it closely.

\section{Top candidates for the heated swings}
\label{sec:top}

Tables~\ref{tab:topTSA} and \ref{tab:topPTSA} list the 20 MOFs with the largest working capacity among those that regenerate below the amine benchmark in TSA and PTSA, respectively. Fig.~6 of the main text shows the first nine of each, with Mg-MOF-74 (OPAGIX), the top TSA candidate, in place of UMCM-313 (EMIZAD) for PTSA. The two lists differ markedly. The TSA candidates are steep (11 of 20) and gradual Langmuir MOFs with $Q^0_\mrm{st}$ between 31 and 60\,kJ\,mol$^{-1}$, mostly on Fe, Mg and Zn nodes. The PTSA candidates are instead mostly S-shaped and two-step MOFs (17 of 20), which bind CO$_2$ weakly, with $Q^0_\mrm{st}$ below 40\,kJ\,mol$^{-1}$ and, for all but one, $K_\mrm{H}$ below 100\,mmol\,g$^{-1}$\,bar$^{-1}$, built mainly on Cu paddlewheels and Zn clusters. No MOF appears in both lists.

We have also repeated the screen with uMOF-MH, the second most accurate potential of our benchmark (Table~\ref{tab:male_mof}), with its isotherms at 298 and 373\,K and the PBE-D3 functional. Most of the UMA-m candidates perform well with uMOF-MH too. Of the 19 TSA and 18 PTSA candidates with a uMOF-MH PES, 18 and all 18, respectively, remain among the best 10\,\% of their swing by working capacity and regenerate below the amine benchmark. The top of the uMOF-MH ranking, however, is different. Only two of the TSA candidates, Mg-MOF-74 and TIXCIO, and four of the PTSA candidates, UTSA-98, UMCM-313, YUZPIU and XAWVUN, are also among its own top 20. Its TSA leaders are mostly narrow-pore Cu, Cd and Mn frameworks for which uMOF-MH predicts near-complete pore filling at 0.1\,bar, whereas UMA-m and UFF predict little uptake. In these frameworks, uMOF-MH binds CO$_2$ more strongly than UMA-m by 0.1--0.2\,eV throughout the pore, an offset that the MACE-MH-1 head of the same model largely does not show. Whether this reflects r$^2$SCAN-D4 itself or the fine-tuning remains to be seen. We therefore report the candidates of UMA-m, the most accurate potential of our benchmark.

\begin{table*}[h]
\small
\centering
\setlength{\tabcolsep}{3pt}
\renewcommand{\arraystretch}{1.1}
\caption{\tbf{Top 20 candidates for TSA}, adsorbing at 0.1\,bar and 298\,K and desorbing at 373\,K. Ranked by the working capacity $\Delta n$, among the MOFs in the best 10\,\% of the swing by $\Delta n$ that regenerate below the amine benchmark of 3.6\,GJ\,t$^{-1}$, as in Fig.~6 of the main text. UMA-m external potential and PBE-D3 CO$_2$ functional, with $K_\mrm{H}$ and $Q^0_\mrm{st}$ at 298\,K. Names are those of the CoRE MOF 2025 database. Where it gives none, an analog is the named MOF with the same linkers and topology (MOFid), and a dash marks a MOF with neither. OMS: open metal sites. Topology (RCSR) from the CoRE MOF 2025 database.}
\label{tab:topTSA}
\begin{tabular}{llccccclcc}
\hline
\hline
{\bfseries\boldmath Refcode} & {\bfseries\boldmath Name} & {\bfseries\boldmath Metal} & {\bfseries\boldmath OMS} & {\bfseries\boldmath Topology} & {\bfseries\boldmath \begin{tabular}[c]{@{}c@{}}$K_\mrm{H}$\\{}[mmol\,g$^{-1}$\,bar$^{-1}$]\end{tabular}} & {\bfseries\boldmath \begin{tabular}[c]{@{}c@{}}$Q^0_\mrm{st}$\\{}[kJ\,mol$^{-1}$]\end{tabular}} & {\bfseries\boldmath Class} & {\bfseries\boldmath \begin{tabular}[c]{@{}c@{}}$\Delta n$\\{}[mmol\,g$^{-1}$]\end{tabular}} & {\bfseries\boldmath \begin{tabular}[c]{@{}c@{}}$E_\mrm{regen}$\\{}[GJ\,t$^{-1}$]\end{tabular}} \\
\hline
OPAGIX & Mg-MOF-74 & Mg & yes & -- & $7.4\times10^{2}$ & 43.6 & gradual & 11.11 & 1.07 \\
FAXTAC & -- & Mn & yes & hcb & $5.1\times10^{4}$ & 60.0 & steep & 8.00 & 1.18 \\
DAXHIV & -- & Fe & yes & -- & $6.9\times10^{1}$ & 33.2 & gradual & 7.46 & 0.92 \\
DAXHUH & -- & Fe & yes & -- & $6.2\times10^{1}$ & 32.6 & gradual & 7.01 & 0.93 \\
DUHSIL & HPU-20 analog & Ni & yes & -- & $6.3\times10^{2}$ & 42.8 & gradual & 6.92 & 0.99 \\
TIXCIO & -- & Cd & no & pyr & $7.2\times10^{1}$ & 30.9 & gradual & 6.90 & 0.91 \\
DIWNAA & -- & Mg & yes & twt & $3.9\times10^{2}$ & 39.8 & steep & 6.89 & 1.05 \\
DAXHIV01 & -- & Fe & yes & -- & $5.9\times10^{1}$ & 32.5 & gradual & 6.73 & 0.93 \\
SUPMAU & -- & Co & yes & -- & $5.2\times10^{1}$ & 32.4 & gradual & 6.64 & 0.93 \\
XESKAJ & -- & Mg & yes & -- & $9.7\times10^{2}$ & 46.2 & steep & 6.62 & 1.31 \\
XAKGOJ & HKUST-1 analog & Zn & yes & tbo & $8.5\times10^{2}$ & 42.5 & steep & 6.41 & 1.16 \\
EQUYOH & -- & Zn & yes & -- & $8.0\times10^{2}$ & 43.5 & steep & 6.33 & 1.10 \\
HUJXUI & -- & Zn & yes & -- & $9.1\times10^{2}$ & 44.2 & steep & 6.28 & 1.11 \\
DAXHUH01 & -- & Fe & yes & -- & $5.6\times10^{1}$ & 32.7 & gradual & 6.16 & 0.94 \\
BELYIC & -- & Mg & yes & dia & $1.5\times10^{4}$ & 54.9 & steep & 6.12 & 1.58 \\
YOZBAS & -- & Ag & no & mtn & $6.1\times10^{4}$ & 51.7 & steep & 6.09 & 1.06 \\
KEFBOO & -- & Zn & yes & nbo & $4.6\times10^{2}$ & 41.8 & steep & 6.03 & 1.14 \\
SEVWEY & CHD-1 & Cd & yes & sra & $4.5\times10^{4}$ & 56.5 & steep & 5.99 & 1.07 \\
APAZOI & -- & La & yes & -- & $1.4\times10^{2}$ & 36.0 & gradual & 5.79 & 1.03 \\
WOJZAY & NNU-21 analog & Ce & yes & -- & $6.2\times10^{2}$ & 41.4 & steep & 5.77 & 1.14 \\
\hline
\hline
\end{tabular}
\end{table*}

\begin{table*}[h]
\small
\centering
\setlength{\tabcolsep}{3pt}
\renewcommand{\arraystretch}{1.1}
\caption{\tbf{Top 20 candidates for PTSA}, adsorbing at 10\,bar and 298\,K and desorbing at 0.1\,bar and 373\,K. Ranked by the working capacity $\Delta n$, among the MOFs in the best 10\,\% of the swing by $\Delta n$ that regenerate below the amine benchmark of 3.6\,GJ\,t$^{-1}$, as in Fig.~6 of the main text. UMA-m external potential and PBE-D3 CO$_2$ functional, with $K_\mrm{H}$ and $Q^0_\mrm{st}$ at 298\,K. Names are those of the CoRE MOF 2025 database. Where it gives none, an analog is the named MOF with the same linkers and topology (MOFid), and a dash marks a MOF with neither. OMS: open metal sites. Topology (RCSR) from the CoRE MOF 2025 database.}
\label{tab:topPTSA}
\begin{tabular}{llccccclcc}
\hline
\hline
{\bfseries\boldmath Refcode} & {\bfseries\boldmath Name} & {\bfseries\boldmath Metal} & {\bfseries\boldmath OMS} & {\bfseries\boldmath Topology} & {\bfseries\boldmath \begin{tabular}[c]{@{}c@{}}$K_\mrm{H}$\\{}[mmol\,g$^{-1}$\,bar$^{-1}$]\end{tabular}} & {\bfseries\boldmath \begin{tabular}[c]{@{}c@{}}$Q^0_\mrm{st}$\\{}[kJ\,mol$^{-1}$]\end{tabular}} & {\bfseries\boldmath Class} & {\bfseries\boldmath \begin{tabular}[c]{@{}c@{}}$\Delta n$\\{}[mmol\,g$^{-1}$]\end{tabular}} & {\bfseries\boldmath \begin{tabular}[c]{@{}c@{}}$E_\mrm{regen}$\\{}[GJ\,t$^{-1}$]\end{tabular}} \\
\hline
FEBXIV & CMOF-2 & Cu & yes & wbl & $2.3\times10^{0}$ & 15.9 & S-shaped & 47.89 & 0.41 \\
VAGMEX & SNU-30SC & Zn & yes & lvt & $1.1\times10^{2}$ & 38.7 & two-step & 42.76 & 0.50 \\
XAFFER & DUT-11 & Zn & yes & pts & $6.0\times10^{1}$ & 36.6 & two-step & 39.88 & 0.49 \\
GATMOH & UTSA-98 & Cu & yes & lvt & $1.5\times10^{0}$ & 13.7 & S-shaped & 39.77 & 0.41 \\
EMIZAD & UMCM-313 & Zr & yes & csq & $4.4\times10^{0}$ & 19.8 & S-shaped & 38.50 & 0.46 \\
BAZFUF01 & MOF-14 & Cu & yes & pto & $1.4\times10^{0}$ & 13.5 & S-shaped & 37.89 & 0.41 \\
MEVMEI & FJIH-U1 & U & yes & tbo & $2.8\times10^{0}$ & 20.2 & S-shaped & 37.48 & 0.44 \\
FIFGEI & CPM-30 & In & yes & acs & $1.7\times10^{1}$ & 31.6 & S-shaped & 36.24 & 0.47 \\
XOVPUU & -- & Cu & yes & pts & $2.2\times10^{0}$ & 16.8 & S-shaped & 36.03 & 0.46 \\
XAFFOB & DUT-10(Cu) & Cu & yes & lvt & $1.5\times10^{0}$ & 14.5 & S-shaped & 35.14 & 0.41 \\
YUZPIU & -- & Zn & yes & nts & $5.1\times10^{1}$ & 39.1 & gradual & 34.97 & 0.50 \\
ALULEZ & -- & Ce & no & pcu & $2.3\times10^{0}$ & 18.6 & S-shaped & 34.80 & 0.45 \\
NIBJAK & MOF-HTB' & Cu & yes & tbo & $1.4\times10^{0}$ & 12.7 & S-shaped & 34.40 & 0.33 \\
QOTKOC04 & PCN-167(In) & In & yes & stp & $8.6\times10^{0}$ & 28.8 & gradual & 33.60 & 0.44 \\
EDUVOO & IRMOF-14 & Zn & no & pcu & $1.3\times10^{0}$ & 14.3 & S-shaped & 32.53 & 0.40 \\
FUYCIN & -- & V & no & srs & $1.6\times10^{0}$ & 14.2 & S-shaped & 32.36 & 0.44 \\
BERZAD & NU-1900 & Zr & yes & bcu & $1.2\times10^{1}$ & 28.5 & gradual & 30.99 & 0.52 \\
XEBHOC & -- & Cu & yes & lvt & $1.7\times10^{0}$ & 15.4 & S-shaped & 29.63 & 0.46 \\
XAWVUN & -- & Cu & yes & pts & $1.6\times10^{0}$ & 15.1 & S-shaped & 29.31 & 0.46 \\
TUHDUY & -- & Cu & yes & pts & $1.9\times10^{0}$ & 16.1 & S-shaped & 28.59 & 0.45 \\
\hline
\hline
\end{tabular}
\end{table*}

\clearpage                                   
\bibliography{references}